\documentclass[12pt]{article}
\catcode`\@=11
\def\marginnote#1{}
\def\draftlabel#1{{\@bsphack\if@filesw {\let\thepage\relax
   \xdef\@gtempa{\write\@auxout{\string
      \newlabel{#1}{{\@currentlabel}{\thepage}}}}}\@gtempa
   \if@nobreak \ifvmode\nobreak\fi\fi\fi\@esphack}
        \gdef\@eqnlabel{#1}}
\def\@eqnlabel{}
\def\@vacuum{}
\def\draftmarginnote#1{\marginpar{\raggedright\scriptsize\tt#1}}
\def\draft{\oddsidemargin -.5truein
        \def\@oddfoot{\sl preliminary draft \hfil
        \rm\thepage\hfil\sl\today\quad\mil922 itarytime}
        \let\@evenfoot\@oddfoot \overfullrule 3pt
        \let\label=\draftlabel
        \let\marginnote=\draftmarginnote
   \def\@eqnnum{(\theequation)\rlap{\kern\marginparsep\tt\@eqnlabel}%
\global\let\@eqnlabel\@vacuum}  }

\def\preprint{\twocolumn\sloppy\flushbottom\parindent 1em
        \leftmargini 2em\leftmarginv .5em\leftmarginvi .5em
        \oddsidemargin -.5in    \evensidemargin -.5in
        \columnsep 15mm \footheight 0pt
        \textwidth 250mmin      \topmargin  -.4in
        \headheight 12pt \topskip .4in
        \textheight 175mm
        \footskip 0pt
        \def\@oddhead{\thepage\hfil\addtocounter{page}{1}\thepage}
        \let\@evenhead\@oddhead \def\@oddfoot{} \def\@evenfoot{} }

\def\titlepage{\@restonecolfalse\if@twocolumn\@restonecoltrue\onecolumn
     \else \newpage \fi \thispagestyle{empty}\c@page\z@
        \def\thefootnote{\fnsymbol{footnote}} }

\def\endtitlepage{\if@restonecol\twocolumn \else  \fi
        \def\thefootnote{\arabic{footnote}} \setcounter{footnote}{0}}

\catcode`@=12
\relax

\def\bea{\begin{array}}
\def\bem{\begin{displaymath}}
\def\beq{\begin{equation}}

\def\eea{\end{array}}
\def\eem{\end{displaymath}}
\def\eeq{\end{equation}}

\def\s2w{\sin^2 \theta_W}
\def\Tr{\mathop{\rm Tr}}

\relax
\def\be{\begin{equation}}
\def\ee{\end{equation}}
\def\ba{\begin{eqnarray}}
\def\ea{\end{eqnarray}}

\def\w{\wedge}
\def\d{{\rm d}}

\def\k{\kappa}
\def\r{\rho}
\def\a{\alpha}

\def\b{\beta}

\def\g{\gamma}
\def\G{\Gamma}

\def\D{\Delta}
\def\e{\epsilon}

\def\m{\mu}
\def\n{\nu}

\def\l{\lambda}

\def\s{\sigma}

\def\cA{{\mathcal A}}
\def\cF{{\mathcal F}}

\def\Ct{{\widetilde C}}
\def\Gt{{\widetilde G}}

\def\Rt{{\widetilde R}}

\def\et{{\widetilde\epsilon}}

\def\St{{\widetilde S}}

\def\IR{\relax{\rm I\kern-.18em R}}

\def\be{\begin{equation}}
\def\ee{\end{equation}}
\def\ba{\begin{eqnarray}}
\def\ea{\end{eqnarray}}

\def\tr{\,{\rm tr}\,}

\def\a{\alpha}
\def\b{\beta}
\def\g{\gamma}
\def\G{\Gamma}
\def\d{{\rm d}}

\def\e{\epsilon}

\def\m{\mu}
\def\n{\nu}
\def\r{\rho}
\def\l{\lambda}

\def\k{\kappa}
\def\s{\sigma}
\def\o{\omega}

\def\vf{\varphi}

\def\ks{{k \kern-.5em /}}
\def\es{{\e \kern-.4em /}}
\def\ds{{\partial \kern-.5em /}}
\def\Ds{{D \kern-.7em /}}

\def\R{{\mathcal R}}

\def\rg{\sqrt{\vert g\vert}}

\def\inv{^{\raise.15ex\hbox{${\scriptscriptstyle -}$}\kern-.05em 1}}

\def\st{{}^*}

\renewcommand{\theequation}{\thesection.\arabic{equation}}
\begin{document}
\topmargin-2.4cm
%
%
%
%
\begin{titlepage}
\begin{flushright}
LPTENS-04/36\\ hep-th/0409028 \\ August 2004
\end{flushright}
\vskip 3.5cm

\begin{center}{\Large\bf 
Anomaly cancellations on lower-dimensional}\\
\vspace{2mm} {\Large\bf 
hypersurfaces by inflow from the bulk} 

\vskip 1.5cm {\bf Adel Bilal}

\vskip.3cm
Laboratoire de Physique Th\'eorique,
\'Ecole Normale Sup\'erieure\\
CNRS\\
24 rue Lhomond, 75231 Paris Cedex 05, France

\end{center}
\vskip .5cm

\begin{center}
{\bf Abstract}
\end{center}
\begin{quote}
Lower-dimensional (hyper)surfaces that can carry gauge or
gauge/gravitational\break\hfill 
anomalies occur in many areas of physics:
one-plus-one-dimensional boundaries or two-dimensional 
defect surfaces in condensed 
matter systems, four-dimensional brane-worlds in 
higher-dimensional cosmologies or various branes and orbifold 
planes in string or M-theory. In all cases we may have (quantum) 
anomalies localized on these hypersurfaces that are only cancelled 
by ``anomaly inflow'' from certain topological interactions in 
the bulk. Proper cancellation between these anomaly contributions 
of different origin requires a careful treatment of factors 
and signs. We review in some detail how these contributions occur 
and discuss applications 
in condensed matter (Quantum Hall Effect) and M-theory (five-branes 
and orbifold planes).
\end{quote}
\end{titlepage}
\newpage
\tableofcontents
\setcounter{footnote}{0}
\setcounter{page}{0}
\setlength{\baselineskip}{.7cm}
\newpage
\noindent
{\sl This work is dedicated to the memory of 
my friend and collaborator Ian Kogan. 
His interest in physics 
spanned almost all of theoretical physics. I hope the present 
contribution goes a little bit in this direction.}

\section{Introduction\label{Intro}}
\setcounter{equation}{0}

Quantum field theories involving chiral fields coupled to
gauge fields and/or gravity may have anomalies. These 
anomalies are a breakdown of gauge, or local Lorentz or 
diffeomorphism invariance respectively, 
at the one-loop level. More specifically, 
the (one-loop) quantum effective action lacks these invariances
necessary for renormalizability and unitarity. In a consistent 
theory all these anomalies must cancel.\footnote{\,%
Of course, anomalies of {\it global} symmetries need not 
cancel and may even be welcome as they allow transitions 
otherwise forbidden by the symmetry.}
Absence of anomalies has been 
much used as a criterion for models in high energy physics, from 
the prediction of the charmed quark \cite{GIM} to the choice of 
gauge group of the superstring \cite{GS}.

Pure gravitational anomalies can only occur in $2, 6, 10, \ldots$
dimensions while gauge or mixed gauge-gravitational anomalies are
possible in all {\it even} dimensions\cite{AGW}. On the other hand, 
quantum field theories in odd dimensions cannot be anomalous. 
Nevertheless, there are many interesting {\it odd}-dimensional 
theories that possess {\it even}-dimensional hypersurfaces with 
chiral matter localized on these surfaces. Typically the chirality 
originates either from an orbifold-like projection, or it is the 
property of a given solution (gravity background) with the 
corresponding ``anti-solution'' having opposite chirality.

Standard examples in eleven-dimensional M-theory \cite{BM} 
are the ten-dimensional
orbifold planes arising from the ${\bf Z}_2$-projection in the 
Horava--Witten realization of the $E_8\times E_8$ heterotic 
string \cite{HW} and the (six-dimensional) five-brane carrying a 
chiral tensor multiplet, while the anti five-brane carries the 
same multiplet but of opposite chirality \cite{W5}. One could 
also mention the $G_2$ compactifications with conical singularities
treated as boundaries of the eleven-dimensional space-time \cite{BMG2}.
Other examples are 
$3+1$-dimensional brane-world cosmologies with chiral matter 
in a 5-dimensional supergravity theory. 
A well-known example in condensed matter is the treatment of 
the $1+1$-dimensional chiral edge currents 
in the $2+1$-dimensional Quantum Hall Effect \cite{QHE}.
One might also consider  chiral vortices, 
again in $2+1$ dimensions,
or two-dimensional 
defect surfaces in three-dimensional (Euclidean) systems.

Typically, these even-dimensional chiral ``subsystems'' possess
one-loop anomalies. This does not contradict the fact that the 
original odd-dimen\-sional theory is anomaly-free. Consider for example
the effective action of eleven-dimensional M-theory. When computing 
the functional integral one has to sum the contributions of 
five-branes and of anti-five-branes and, of course, the anomalous 
contributions, being opposite, cancel. However, we rather like to 
think of M-theory within a given background with some five-branes 
in certain places and anti five-branes in others, maybe far away, 
and require {\it local} anomaly cancellation, i.e. on each (anti) 
five-brane separately, rather than just global cancellation. 
Remarkably, such local cancellation is indeed achieved by a 
so-called ``anomaly inflow'' from the bulk;  M-theory has 
eleven-dimensional Chern--Simons like terms that are invariant 
in the bulk but have a anomalous variations on 
five-branes or on boundaries, precisely cancelling the one-loop
anomalies locally \cite{DLM,FHMM}.

In this paper we will explain in some generality such anomaly inflow
from the bulk and how it can and does cancel the gauge and 
gravitational anomalies on the even-dimensional hypersurfaces.
We will discuss why anomaly inflow always originates from 
topological terms (in odd dimensions). Usually, when discussing 
anomaly 
cancellation between different chiral fields one need not be 
very careful about overall common factors. Here, however, we 
want to consider cancellations between anomaly contributions 
of very different origin and special attention has to be paid 
to all factors and signs (see \cite{BM}). 
To this end, we also discuss in some 
detail the continuation between Euclidean and Minkowski signature,
which again sheds some new light on why it must be the topological 
terms that lead to anomaly inflow.

In the next section, we begin by
a general discussion of anomalies and anomaly inflow from the bulk, 
spending some time and space on the subtle continuation between 
Euclidean and Minkowski signature. Part of this section is just 
a recollection of standard results on one-loop anomalies \cite{AGG} 
with special attention to conventions and signs. We explain how
anomaly inflow uses the descent equations and why it necessarily 
originates from a manifold of higher dimension than the one on 
which the anomalous theory lives. Section 3 describes an 
elementary application to the (integer) Quantum Hall Effect 
where the effective bulk theory is a Chern--Simons theory and the 
boundary degrees of freedom are the chiral edge currents; 
anomaly cancellation by inflow from the bulk correctly explains
the quantized Hall conductance. In Section 4, we
describe two examples of anomaly cancellation by inflow 
in M-theory in quite some detail: 
on five-branes and on the ${\bf Z}_2$-orbifold planes. 
For the five-branes, in order to get all signs and coefficients
consistent, we rederive everything from scratch: the
solution itself, the modified Bianchi identity, the zero-modes
and their chirality, the one-loop anomaly and the 
FHMM \cite{FHMM} mechanism for the inflow. For the 
orbifold planes we insist on the correct normalization 
of the Bianchi identity and describe the modification of 
the Chern--Simons term obtained in \cite{BM} necessary to get the 
correct inflow. Finally, in Section 5, we 
briefly mention analogous cancellations in brane-world scenarios.

\section{One-Loop Anomalies and Anomaly Inflow\label{anomalies}}
\setcounter{equation}{0}

\subsection{Conventions\label{conventions}}

We begin by carefully defining our conventions. They are the same as
in \cite{BM}. In  Minkowskian space we always
use signature $(-,+,\ldots,+)$ and label the coordinates $x^\m$,
$\m=0,\ldots D-1$. We always choose a {\it right}-handed
coordinate system such that
\be\label{d1}
\int \rg\ \d x^0\w\d x^1\w \ldots \w \d x^{D-1}
= + \int\rg\ \d^D x \ge 0  \ .
\ee
(With $x^0$ being time and for even $D$, this is a non-trivial
statement. In particular, for even $D$, if we relabelled time as
$x^0\to x^D$ then $x^1,\ldots x^D$ would be a left-handed
coordinate system!) We define the $\e$-tensor as
\be\label{d2}
\e_{01\ldots (D-1)}
=+\rg \quad \Leftrightarrow \quad \e^{01\ldots (D-1)}
=-{1\over \rg} \ .
\ee
Then \be\label{d3} \d x^{\m_1} \w
\ldots \w \d x^{\m_D} =
 - \e^{\m_1\ldots\m_D}\, \rg\ \d^D x \ .
\ee
A $p$-form $\o$ and its components are related as
\be\label{d4}
\o={1\over p!}\, \o_{\m_1\ldots\m_p}\,
\d x^{\m_1} \w \ldots \w \d x^{\m_p}
\ee
and its dual is
\be\label{d5}
\st\o = {1\over p! (D-p)!}\, \o_{\m_1\ldots\m_p}\
\e^{\m_1\ldots\m_p}_{\phantom{\m_1\ldots\m_p} \m_{p+1}\ldots\m_D}
\ \d x^{\m_{p+1}}\w\ldots\w\d x^{\m_D}\ .
\ee
We have $\st(\st\o)=(-)^{p(D-p)+1}\,\o$ and
\be\label{d6} \o\w \st\o= {1\over p!}\, \o_{\m_1\ldots\m_p}\,
\o^{\m_1\ldots\m_p}\ \rg\ \d^Dx \ . \ee Finally we note that the
components of the $(p+1)$-form $\xi=\d\o$ are given by
\be\label{d8} \xi_{\m_1\ldots\m_{p+1}}= (p+1)\,
\partial_{[\m_1}\o_{\m_2\ldots\m_{p+1}]}
\end{equation}
(where the brackets denote anti-symmetrization with total weight
one) and that the divergence of a $p$-form is expressed as
\be\label{d7} \st\d \st\o={(-)^{D(p-1)+1}\over (p-1)!} \,
\nabla^\n \o_{\n\m_1\ldots\m_{p-1}}\, \d x^{\m_1} \w \ldots \w \d
x^{\m_{p-1}} \ . \ee

We define the curvature 2-form $R^{ab}={1\over 2} R^{ab}_{\ \ \
\n\s}\, \d x^\n\w\d x^\s$ in terms of the spin-connection
$\o^{ab}$ as $R^{ab}=\d\o^{ab} + \o^a_{\ c}\w \o^{cb}$. Here
$a,b,c = 0, \ldots D-1$ are ``flat'' indices, related to the
``curved'' ones by the $D$-bein $e^a_\m$. The torsion is $T^a=\d
e^a+ \o^a_{\ b}\w e^b$. The Riemann tensor $R^{\m\r}_{\ \ \ \n\s}$
is related to the curvature 2-form via $R^{ab}_{\ \ \ \n\s}=
e^a_\m e^b_\r R^{\m\r}_{\ \ \ \n\s}$, and the Ricci tensor is
$\R^\m_{\ \ \n} = R^{\m\r}_{\ \ \ \n\r}$ while the Ricci scalar
$\R$ is given by $\R=\R^\m_{\ \ \m}$. With this sign convention,
(space-like) spheres have $\R>0$.

For gauge theory, the gauge
fields, field strength and gauge variation are given by
\ba\label{q1}
 \nonumber A&=& A_\m\,\d z^\m\ ,
\quad A_\m=A_\m^\a\, \l^\a\ ,
\quad (\l^\a)^\dag = - \l^\a \ , \\[1mm]
F&=& \d A + A^2 \equiv  \d A + A\w A \ , \quad \delta_v A= {\rm D} v= \d v+[A,v] \ .
\ea
Thus $F$ is anti-hermitian and differs by an $i$ from a
hermitian field strength used by certain authors.\footnote{\,%
For $U(1)$-gauge theories, the usual definition of the
covariant derivative is $\partial_\m+iq{\mathcal A}_\m$, with $q$
being the charge, and
hence $A\simeq i q {\mathcal A}$ and $F\simeq iq{\mathcal F}$
where ${\mathcal F}=\d {\mathcal A}$.
}
For gravity, one
considers the spin connection $\o^a_{\ b}$ as an $SO(2n)$-matrix
valued 1-form. Similarly, the parameters $\e^a_{\ b}$ of local
Lorentz transformations (with $\e^{ab}=-\e^{ba}$) are considered
as an $SO(2n)$-matrix. Then
\be\label{q2}
R=\d\o+\o^2\ , \quad
\delta_\e e^a=-\e^a_{\ b}e^b \ , \quad
\delta_\e \o={\rm D}\e=\d\e + [\o,\e] \ .
\ee
For spin-${1\over 2}$ fermions the relevant Dirac
operator is  ($E_a^\m$ is the inverse $2n$-bein)
\be\label{qq3}
\Ds=E_a^\m \g^a \left( \partial_\m +A_\m
+{1\over 4} \o_{cd,\m} \g^{cd}\right) \ , \quad \g^{cd}={1\over 2}
[\g^c,\g^d] \ .
\ee

\subsection{Continuation Between Minkowski and Euclidean 
Signature\label{continuation}}

We now turn to the continuation to Euclidean signature. 
While the Minkowskian functional integral   contains
$e^{iS_{\rm M}}$, the Euclidean one contains $e^{-S_{\rm E}}$.
This implies 
\be\label{d15} 
S_{\rm M}=i\, S_{\rm E} \quad , \quad
x^0=-i\, x^0_{\rm E} \ . 
\ee 
However, for a Euclidean manifold
$M_{\rm E}$ it is natural to index the coordinates from 1 to $D$,
not from $0$ to $D-1$. One could, of course, simply write
$ix^0=x^0_{\rm E}\equiv x^D_{\rm E}$. The problem then is for even
$D=2n$ that $\d x^0_{\rm E}\w \d x^1\w\ldots\d x^{2n-1} = -\ \d
x^1\w\ldots\d x^{2n-1} \w \d x^{2n}_{\rm E}$ and if $(x^0_E,\ldots
x^{2n-1})$ was a right-handed coordinate system then $(x^1,\ldots
x^{2n}_E)$ is a left-handed one. This problem is solved by
shifting the indices of the coordinates as 
\be\label{d16} 
i\, x^0 = x^0_{\rm E} = z^1 \ , \quad x^1=z^2 \ , 
\quad \ldots \ , \quad
x^{D-1} = z^D \ . 
\ee 
This is equivalent to a specific choice of
an orientation on the Euclidean manifold $M_{\rm E}$. In
particular, we impose 
\be\label{d16a} 
\int \sqrt{g}\, \d z^1\w
\ldots\w\d z^D = + \int \sqrt{g}\, \d^D z \ge 0 \ . 
\ee 
Then, of
course, for any tensor we similarly shift the indices, e.g.
$C_{157}=C^{\rm E}_{268}$ and $C_{034}=i\, C^{\rm E}_{145}$. We
have $G_{\m\n\r\s}\, G^{\m\n\r\s} = G^{\rm E}_{jklm}\, G_{\rm
E}^{jklm}$ as usual, and for a $p$-form
\be\label{d17}
\o={1\over p!}\, \o_{\m_1\ldots\m_p}\,
 \d x^{\m_1}\w\ldots\w\d x^{\m_p}
={1\over p!}\, \o^{\rm E}_{j_1\ldots j_p}\,
 \d z^{j_1}\w\ldots\w\d z^{j_p} =\o^{\rm E} \ .
\ee
In particular, we have for $p=D$
\be\label{d18}
\int_{M_{\rm M}} \o = \int_{M_{\rm E}} \o^{\rm E}\ ,
\ee
which will be most important below.
 Finally, note that the Minkowski relations
(\ref{d2}) and (\ref{d3}) become
\be\label{d19}
\d z^{j_1}\w \ldots\w\d z^{j_D}=
+\, \e_{\rm E}^{j_1\ldots j_D}\, \sqrt{g}\, \d^D z
\quad {\rm with} \quad
\e_{\rm E}^{1\ldots D}={1\over \sqrt{g}} \ .
\ee
The dual of a $p$-form $\o^{\rm E}$ is defined as in (\ref{d5})
but using $\e_{\rm E}$. It then follows that
$\st(\st\o_{\rm E})=(-)^{p(D-p)}\, \o_{\rm E}$ (with an additional
minus sign with respect to the Minkowski relation) and, as in
the Minkowskian case,
$\o_{\rm E}\w \st\o_{\rm E}
= {1\over p!}\, \o^{\rm E}_{j_1\ldots j_p}\,
\o_{\rm E}^{j_1\ldots j_p}\, \sqrt{g}\ \d^D z$.

It follows from the preceding discussion that the Euclidean 
action is not always real, not even its bosonic part. The 
original (real) Minkowskian action can contain two types of 
(locally) Lorentz invariant terms, terms that involve the 
metric like such as
\be\label{dmetricM}
S_{\rm M}^{(1)} = {1\over 2} \int \tr F \w {}^*F
 = {1\over 4} \int \tr F_{\m\n} F^{\m\n} \rg\, \d^{D}x
\ee
and topological (Chern--Simons type) terms that do not 
involve the metric such as (if $D$ is odd)
\be\label{dtopolM}
S_{\rm M}^{(2)} = \int \tr A\w F\w \ldots \w F \ .
\ee
It follows from (\ref{d15}), (\ref{d16}) and (\ref{d18}) that the 
Euclidean continuations of these two terms are
\be\label{dmetricE}
S_{\rm E}^{(1)} = -{1\over 2} \int \tr F_{\rm E} \w {}^*F_{\rm E}
 = - {1\over 4} \int \tr F^{\rm E}_{\m\n} F^{\m\n}_{\rm E} 
\sqrt{g}\, \d^{D}z
\ee
and 
\be\label{dtopolE}
S_{\rm E}^{(2)} = 
- i \int \tr A_{\rm E}\w F_{\rm E}\w \ldots \w F_{\rm E} \ .
\ee
Hence the imaginary part of the Euclidean bosonic action is given 
by the topological terms. Note that from now on  we will not
write the wedge products explicitly, but $\tr A F^2$ will be short-hand for 
$\tr A\w F\w F$, etc.

There is a further subtlety that needs to be settled when discussing 
the relation between the Minkowskian and Euclidean forms of the 
anomalies. One has to know how
the chirality matrix $\g$ is continued from the Euclidean to the
Minkowskian and vice versa. This will be relevant for the
$2n$-dimensional submanifolds. The continuation of the
$\g$-matrices is dictated by the continuation of the coordinates
we have adopted (cf (\ref{d16})):
\be\label{d30}
i\, \g^0_{\rm M}=
\g^1_{\rm E} \ , \quad \g^1_{\rm M}= \g^2_{\rm E} \ ,
\quad \ldots
\quad \g^{2n-1}_{\rm M}= \g^{2n}_{\rm E} \ .
\ee
In accordance
with Ref. \cite{AGG} we define the Minkowskian and Euclidean
chirality matrices $\g_{\rm M}$ and $\g_{\rm E}$ in $2n$
dimensions as
\be\label{d31}
\g_{\rm M}=i^{n-1} \g^0_{\rm M}\ldots \g^{2n-1}_{\rm M}
\quad , \quad
\g_{\rm E} = i^n \g^1_{\rm E} \ldots  \g^{2n}_{\rm E} \ .
\ee
Both $\g_{\rm M}$ and $\g_{\rm E}$ are hermitian.
Taking into account (\ref{d30}) this leads to
\be\label{d32}
\g_{\rm M}= -  \g_{\rm E} \ ,
\ee
i.e. what we call
positive chirality in Minkowski space is called negative chirality
in Euclidean space and vice versa. This relative minus sign is
somewhat unfortunate, but it is necessary to define self-dual
$n$-forms from a pair of positive chirality spinors, both in 
Minkowskian space (with our convention for the $\e$-tensor) and in
Euclidean space (with the conventions of
\cite{AGG}).\,\footnote{\,Since we will take \cite{AGG} as the
standard reference for computing anomalies in  Euclidean space, we
certainly want to use the same convention for $\g_{\rm E}$. On the
other hand, we have somewhat more freedom to choose a sign
convention for $\g_{\rm M}$. The definition (\ref{d31}) of
$\g_{\rm M}$ has the further advantage that in $D=10$, $\g_{\rm
M}= \g^0_{\rm M}\ldots \g^9_{\rm M}$ which is the usual convention
used in string theory \cite{GSW}. Our $\g_{\rm M}$ also agrees
with the definition of \cite{POL} in $D=2$, 6 and 10 (but
differs from it by a sign in $D=4$ and 8).}

Indeed, as is well-known, in $2n=4k+2$ dimensions, from a pair of
spinors of the same chirality one can always construct the
components of an $n$-form $H$ by sandwiching $n$ (different)
$\g$-matrices between the two spinors. In Minkowskian space we call
such an $n$-form $H^{\rm M}$ self-dual if 
\be\label{d33} 
H^{\rm M}_{\m_1\ldots \m_n} = +\, {1\over n!}\, \e_{\m_1 \ldots \m_{2n}}
H_{\rm M}^{\m_{n+1} \ldots \m_{2n}} 
\ee 
(with $\e$ given by
(\ref{d2})) and it is obtained from 2 spinors $\psi_I$ ($I=1,2$)
satisfying $\g_{\rm M}\,\psi_I=+\psi_I$. In  Euclidean space $H^{\rm
E}$ is called self-dual if (cf \cite{AGG}) 
\be\label{d34} 
H^{\rm E}_{j_1\ldots j_n} = +\, {i\over n!}\, \e^{\rm E}_{j_1 \ldots
j_{2n}} H_{\rm E}^{j_{n+1} \ldots j_{2n}} 
\ee 
(with $\e^{\rm E}$
given by (\ref{d19})) and it is obtained from 2 spinors $\chi_I$
($I=1,2$) satisfying $\g_{\rm E}\,\chi_I=+\chi_I$. With these
conventions a self-dual $n$-form in Minkowski space continues to
an {\it anti}-self-dual $n$-form in Euclidean space, and vice
versa, consistent with the fact that positive chirality in
Minkowski space continues to negative chirality in Euclidean
space. The situation is summarized in Table \ref{table2} where
each of the four entries corresponds to any of the 3 others.

\begin{table}[h]
\begin{center}
\begin{tabular}{|c||c|c|}\hline
        & Minkowskian            & Euclidean    \\
\hline
\hline
spinors & positive chirality &  negative chirality   \\
\hline
$n$-form& self-dual          & anti-self-dual    \\
\hline
\end{tabular}
\caption{\it Correspondences between the (anti-) self-duality of
$n$-forms in $2n=4k+2$ dimensions and the chirality of the
corresponding pair of spinors are given, as well as their
Euclidean, resp. Minkowskian continuations.} \label{table2}
\end{center}
\end{table}

As we will recall below, the anomalies are given by topological terms
$\int_{M_{\rm M}^{2n}} D_{\rm M}^{(2n)}$ whose continuation is simply 
$\int_{M_{\rm E}^{2n}} D_{\rm E}^{(2n)}$ (cf Eq.\,(\ref{d18})) where
$ D_{\rm M}^{(2n)}$ is the anomaly expression obtained by 
continuation from $ D_{\rm E}^{(2n)}$ with {\it the  chiralities 
corresponding as discussed above}. One also has to remember that 
the continuation of the effective action $\G$ includes an extra 
factor $i$ according to Eq.\,(\ref{d15}). In conclusion, 
the anomaly of a positive chirality spinor
(or a self-dual $n$-form) in
Minkowski space is given by $\delta\Gamma_M=\int_{M_M^{2n}}\hat
I_{2n}^1$ if  in Euclidean space the anomaly of
a negative chirality spinor (or an
anti-self-dual $n$-form) is given by 
$\delta\Gamma_E= - i \int_{M_E^{2n}}\hat
I_{2n}^1$. This will be discussed in more detail in the next subsection.

\subsection{The One-Loop Anomalies\label{oneloop}}

This subsection is a summary of the results of \cite{AGG} where
the anomalies for various chiral fields in  Euclidean space  were 
related to index theory. This whole subsection will be in Euclidean 
space of even dimension $2n$. We first give the different relevant 
indices. The simplest index is that of a positive chirality 
spin-${1\over 2}$ field. Here positive chirality means positive 
Euclidean chirality as defined above. For spin-${1\over 2}$ 
fermions the relevant Euclidean Dirac operator is (cf. (\ref{qq3}))
\be\label{q3}
\Ds=E_a^j \g^a \Big( \partial_j +A_j
+{1\over 4}\, \o_{cd,j} \g^{cd}\Big) \, , \quad \g^{cd}={1\over 2}\,
[\g^c,\g^d] \ .
\ee
Define $\Ds_{1\over2}=\Ds\ {{1+\g}\over2}$
and the index as
\begin{eqnarray}\label{indexdef}
{\rm ind}\big(i\Ds_{1\over2}\big)&=&\mbox{ number of zero modes of}\
i\Ds_{1\over2} \nonumber\\
&&\hskip-1.5mm -\, \mbox{number of zero modes of}\
\big(i\Ds_{1\over2}\big)^\dagger\ .
\end{eqnarray}
Then by the Atiyah--Singer index theorem
\be
{\rm ind}\big(i\Ds_{1\over2}\big)=\int_{M_{2n}}[\hat A(M_{2n})\ {\rm
ch}(F)]_{2n}\,,
\ee
where ${\rm ch}(F)=\tr\exp\left({i\over 2\pi}F\right)$ is the
Chern character and $\hat A(M_{2n})$ is the Dirac genus of the
manifold, given below. The subscript $2n$ indicates to pick only
the part which is a $2n$-form. Note that if the gauge group is
$\prod_{k} G_k$, then ${\rm ch}(F)$ is replaced by
$\prod_k {\rm ch}(F_k)$.

Another important index is that of a positive chirality
spin-${3\over2}$ field. Such a field is obtained from a positive
chirality spin-${1\over2}$ field with an extra vector index by
subtracting the spin-${1\over2}$ part. An extra vector index leads
to an additional factor for the index density,
\begin{equation}
\tr \exp\Big({i\over2\pi}\,{1\over2}R_{ab}T^{ab}\Big)=\tr
\exp \Big({i\over2\pi}R \Big)\,,
\end{equation}
since the vector representation is
$(T^{ab})_{cd}=\delta^a_c\delta^b_d-\delta^a_d\delta^b_c$. Hence
\begin{equation}
{\rm ind}\big(iD_{3\over2}\big)=\int_{M_{2n}}\left[\hat A(M_{2n})\Big(\tr
\exp\Big({i\over2\pi}R \Big)-1\Big)\ {\rm ch}(F)\right]_{2n}.
\end{equation}

The third type of field which leads to anomalies is a self-dual or
anti-self-dual $n$-form $H$ in $2n=4k+2$ dimensions.
Such antisymmetric tensor fields
carry no charge w.r.t. the gauge group. As discussed above, 
a self-dual tensor can be constructed from a
pair of positive chirality spinors. Correspondingly, the index is
$\hat A(M_{2n})$ multiplied by $\tr
\exp\left({i\over2\pi}{1\over2}R_{ab}T^{ab}\right)$, where
$T^{ab}={1\over2}\gamma^{ab}$ as appropriate for the
spin-${1\over2}$ representation. Note that the trace over the
spinor representation gives a factor $2^n$ in $2n$ dimensions.
There is also an additional factor ${1\over2}$ from the chirality
projector of this second spinor and another factor ${1\over2}$
from a reality constraint ($H$ is real),
\begin{equation}
{\rm ind}(iD_A)
={1\over4}\int_{M_{2n}} \left[\hat A(M_{2n})
\tr \exp\Big({i\over2\pi}{1\over4}R_{ab}\g^{ab}\Big)
\right]_{2n}
={1\over4}\int_{M_{2n}}[L(M)]_{2n}\,.
\label{indDA}
\end{equation}
$L(M)$ is called the Hirzebruch polynomial, and the subscript on
$D_A$ stands for ``antisymmetric tensor".
(Note that, while $\hat A(M_{2n})\tr
\exp\left({i\over2\pi}{1\over4}R_{ab}\gamma^{ab}\right)$ carries
an overall factor $2^n$,
$L(M_{2n})$ has a factor $2^k$ in front of each
$2k$-form part. It is only for $k=n$ that they coincide.)

Of course, the index of a negative chirality (anti-self-dual)
field is minus that of the corresponding positive chirality
(self-dual) field. Explicitly one has:
\ba
{\rm ch}(F)
&=&
\tr \exp\left({i\over2\pi}F\right)
=\tr{\bf 1}\!+\!{i\over2\pi}\tr F\!+\ldots
+{i^k\over k!(2\pi)^k}\tr F^k\!+\ldots\,,~~~~~~~~
\label{chF}\\[1mm]
\hat A(M_{2n})
&=& 1+{1\over (4\pi)^2}{1\over12}\tr R^2
+{1\over(4\pi)^4}\left[{1\over360}\tr R^4
+{1\over288}(\tr R^2)^2\right]\nonumber\\[1mm]
&&\hskip-2.mm+{1\over(4\pi)^6}\left[{1\over5670}\tr R^6
+{1\over4320}\tr R^4\tr R^2+{1\over10368}(\tr R^2)^3\right]+\ldots\,,
\label{Aroof}
\ea
\vskip -0.1cm
\ba
&&
\hat A(M_{2n})\left(\tr {\rm e}^{{i\over2\pi}R} -1\right)
=(2n-1)+{1\over (4\pi)^2}{2n-25\over12}\tr R^2
\nonumber\\
&&
\hskip 24mm +{1\over(4\pi)^4}\left[{2n+239\over360}\tr R^4
+{2n-49\over288}(\tr R^2)^2\right]
\nonumber\\[1mm]
&&\hskip 24mm +{1\over(4\pi)^6}\left[{2n-505\over5670}\tr R^6
+{2n+215\over4320}\tr R^4\tr R^2
+{2n-73\over10368}(\tr R^2)^3\right]+\ldots\,,
\nonumber\\[-1mm]
\label{ArooftrR}
\ea
\vskip -0.7cm
\ba
L(M_{2n})
&=&1-{1\over (2\pi)^2}{1\over6}\tr R^2
+{1\over(2\pi)^4}\left[-{7\over180}\tr R^4
+{1\over72}(\tr R^2)^2\right]
\nonumber\\[1mm]
&&
\hskip-2.mm
+{1\over(2\pi)^6}\left[-{31\over2835}\tr R^6
+{7\over1080}\tr R^4\tr R^2-{1\over1296}(\tr R^2)^3\right]+\ldots\,.~~~~~~
\label{L(M)}
\ea

To proceed, we need to define exactly what we mean by the anomaly. 
For the time being, we suppose that the classical action is invariant (no
inflow), but that the Euclidean quantum effective action $\G_E[A]$
has an anomalous variation under the gauge transformation
(\ref{q1}) with parameter $v$ of the form
\be
\delta_v\G_E[A]=\int\tr v\, \mathcal{G}(A)\,.
\ee
Local Lorentz anomalies are treated analogously. Note that
\be
\delta_v\G_E[A]=\int (D_\m v)^\a{\delta\G_E[A]\over\delta
A_\m^\a}=-\int v^\a(D_\m J^\m)^\a
\ee
or\,\footnote{\,Note that if $A=A^\a\l^\a$, $B=B^\b\l^\b$ and $\tr
\l^\a\l^\b=-\delta^{\a\b}$ (the $\l^\a$ are anti-hermitian) then
e.g. $\tr AB=-A^\a B^\a$ and ${\delta\over\delta A^\alpha}\int \tr
AB=-B^\a$. Hence one must define ${\delta\over\delta
A}=-\l^\a{\delta\over\delta A^\a}$ so that ${\delta\over\delta
A}\int \tr AB=B$. Another way to see this minus sign in
${\delta\over\delta A }$ is to note that $A^\a=-\tr \l^\a A$.}
\be
\delta_v\G_E[A]=\int \tr D_\m v{\delta\G_E[A]\over\delta
A_\m}=-\int\tr  v D_\m {\delta\G_E[A]\over\delta A_\m}
\ee
so that $\mathcal{G}(A)$ is identified with
$-D_\m{\delta\G_E[A]\over\delta A_\m}$ or $\mathcal{G}(A)^\a$ with
$-(D_\m J^\m)^\a$. To avoid these complications, 
we will simply refer to the anomalous variation of the
effective action, $\delta_v\G_E[A]$ as the
anomaly. So our anomaly is the negative integrated divergence of
the quantum current (multiplied with the variation parameter $v$).

A most important result of \cite{AGG} is the precise relation
between the anomaly in $2n$ dimensions and index theorems in
$2n+2$ dimensions, which for the pure gauge anomaly of a positive
chirality spin-${1\over2}$ field is (Eq.\,(3.35) of \cite{AGG})
\be
\delta_v\G_E^{spin {1\over2}}[A]=+{i^n\over(2\pi)^n(n+1)!}\int
Q_{2n}^1(v,A,F)\,.
\ee
The standard descent equations
$\d Q_{2n}^1=\delta_vQ_{2n+1}$ and $\d Q_{2n+1}=\tr F^{n+1}\ $
relate $Q_{2n}^1$ to the invariant polynomial $\tr F^{n+1}$.
Comparing with (\ref{chF}) we see that the pure gauge anomaly is
thus given by
$\delta_v\G_E^{spin {1\over2}}[A]=
\int I_{2n}^{1,gauge}$ with the descent equations
$\d I_{2n}^{1,gauge}=\delta_v I_{2n+1}^{gauge}$ and
$\d I_{2n+1}^{gauge}=I_{2n+2}^{gauge}$, where
$I_{2n+2}^{gauge}= -2\pi i\, [{\rm ch}(F)]_{2n+2}\,$.
This is immediately generalized to include all gauge and local
Lorentz anomalies due to all three types of chiral fields
\begin{eqnarray}
\delta\G_E[A]&=&\int I_{2n}^1\,,\\[1mm]
\d I_{2n}^1=\delta I_{2n+1}&,&
\d I_{2n+1}=I_{2n+2}\, ,
\label{-2piiindex}
\end{eqnarray}
where $I_{2n+2}$ equals $-2\pi i$ times the relevant index density
appearing in the index theorem in $2n+2$ dimensions (corrected
by a factor of $\left(-{1\over2}\right)$ in the case of the
antisymmetric tensor
field, see below). This shows that the
Euclidean anomaly is purely imaginary. It is thus convenient to
introduce $\hat I$ as $I=-i\ \hat I$ so that\\[-8mm]~
\begin{eqnarray}
\delta\G_E[A]&=&-i\int \hat I_{2n}^1\,,\label{dG=-iIhat}\label{Eanom}\\[1mm]
\d\hat I_{2n}^1=\delta \hat I_{2n+1}&,&
\d\hat I_{2n+1}=\hat I_{2n+2}\, .
\label{-2piindex}
\end{eqnarray}
Explicitly we have (always for positive Euclidean chirality, respectively Euclidean self-dual forms)
\begin{eqnarray}\label{Ihathalf}
\hat I_{2n+2}^{spin{1\over2}}&=&2\pi \left[\hat A(M_{2n})\ {\rm
ch}(F)\right]_{2n+2}\,,\\[1mm]
\hat I_{2n+2}^{spin{3\over2}}&=&2\pi \left[\hat A(M_{2n})\
\left(\tr \exp\left({i\over2\pi}R\right)-1\right)\ {\rm
ch}(F)\right]_{2n+2}\,,\label{Ihat3half}\\[1mm]
\hat I_{2n+2}^{A}&=&2\pi \left[\left(-{1\over2}\right){1\over4}\
L(M_{2n})\right]_{2n+2}.\label{IhatA}
\end{eqnarray}
The last equation contains an extra factor
$\left(-{1\over2}\right)$ with respect to the index (\ref{indDA}).
The minus sign takes into account the Bose rather than Fermi
statistics, and the $1\over2$ corrects the $2^{n+1}$ to $2^n$
which is the appropriate dimension of the spinor representation on
$M_{2n}$ while the index is computed in $2n+2$ dimensions. Note
that in the cases of interest, the spin-${3\over2}$ gravitino is
not charged under the gauge group and in (\ref{Ihat3half}) the
factor of ${\rm ch}(F)$ simply equals 1.

Equations (\ref{dG=-iIhat})-(\ref{IhatA}) together with
(\ref{chF})-(\ref{L(M)}) give explicit expressions for the
anomalous variation of the Euclidean effective action. In the previous
subsection we carefully studied the continuation of topological terms
like $\int \hat I_{2n}^1$ between Minkowski and Euclidean
signature. It follows from equations (\ref{d15}), (\ref{d18})
and (\ref{Eanom}) that
the anomalous variation of the Minkowskian effective action is
given directly by $\hat I_{2n}^1$,
\be\label{Manom}
\delta\G_M=\int_{M^M_{2n}} \hat I_{2n}^1\,.
\ee
However, one has to remember that (with our conventions for
$\gamma_M$) the chiralities in Minkowski space and Euclidean space
are opposite. While $\hat I^1_{2n}$ corresponds to positive
chirality in the Euclidean, it corresponds to negative chirality
in Minkowski space, i.e. Eq.\,(\ref{Manom}) {\it is the anomaly 
for a negative chirality field in Minkowski space.} Obviously, 
the anomaly of a positive  chirality field in Minkowski space 
is just the opposite.

To facilitate comparison with references
\cite{GSW} (GSW) and \cite{FLO} (FLO) we note that
\be
I_{\rm GSW}=(2\pi)^n\hat I_{2n+2}\,,\quad I_{\rm FLO}=-\hat
I_{2n+2}\,.
\ee
The flip of sign between $I_{\rm FLO}$ and $\hat I_{2n+2}$ is
such that $\int I^1_{\rm FLO}$ directly gives the variation
of the Minkowskian effective action for positive chirality
spinors in  Minkowskian space (with our definition of $\g_{\rm M}$).

Before we go on, it is perhaps useful to look at an explicit 
example in four dimensions.
Consider the simple case of a spin-${1\over 2}$
fermion of {\it negative} Minkowskian chirality coupled to 
$SU(N)$ gauge fields.
In the Euclidean, this corresponds to positive chirality and hence
the anomalous variation of the Minkowskian effective action is
$\delta\G_M=\int \hat I_4^1$, where $\hat I_4^1$ is related via 
the descent equations to $\hat I_6$ which is obtained from 
(\ref{Ihathalf}) as
\be\label{example1}
\hat I_6= - {i\over 6 (2\pi)^2} \tr\, F^3 \, .
\ee
Note that this is real since by (\ref{q1}) $\tr\, F^3$ is purely imaginary.
Also, there is no mixed gauge-gravitational anomaly since the 
relevant term $\sim \tr R^2 \tr F$ vanishes for $SU(N)$ gauge 
fields. It is only for $U(1)$ gauge fields that one can get a 
mixed gauge-gravitational anomaly in four dimensions.
Using the descent equations one explicitly gets
\be\label{example2}
\delta\G_M=- {i\over 6 (2\pi)^2} 
\int\! \tr v \ \d \Big( A \d A + {1\over 2} A^3 \Big)\, .
\ee
It is important to note that we are only discussing the 
so-called consistent anomaly. Indeed, since our anomaly 
is defined as the variation of the effective action it 
automatically satisfies the Wess-Zumino consistency 
condition \cite{ZW} and
hence is the consistent anomaly. There is also another 
manifestation of the anomaly, the so-called covariant 
anomaly (which in the present example would be 
$- {i\over 2 (2\pi)^2} \int \tr v F^2$). The latter is not 
relevant to us here and we will not discuss it further (see however Ref.
\cite{BZ}).

Finally, it is worth mentioning that the anomalies are
``quantized'' in the following sense: once we have normalized
the gauge and gravitational fields in the usual way (so that
$F=\d A+A^2$ and $R=\d\o+\o^2$) the  anomalies
have no explicit dependence on the gauge or gravitational 
coupling constants. In a given theory, the total anomaly is a sum of
the fixed anomalies $\hat I^{spin{1\over2}}$, 
$\hat I^{spin{3\over2}}$ and $\hat I^A$ with coefficients 
that count the multiplicities of the corresponding fields, 
i.e. are integers. Of course, this came about from the 
relation with index theory, and there was just no place where 
any coupling constants could show up. Another way to see 
this is to recall that the anomalies are one-loop  
contributions to the effective action coming from exponentiating 
determinants. In the loop expansion of the effective action 
only the one-loop term is independent of the coupling constants.
 
\subsection{Anomaly Cancellation by Inflow\label{inflow}}

We have seen that the anomalous variation of the one-loop 
quantum effective action is 
$\delta\G_{\rm E}= - i\int_{M_E^{2n}}\hat I_{2n}^1$ 
in the Euclidean and
$\delta\G_{\rm M}=  \int_{M_M^{2n}} \hat I_{2n}^1$ in the
Minkowskian case. Now, we want to discuss the situation where 
$M_{2n}$ is a $2n$-dimensional submanifold (on which live the chiral 
fields that give rise to the anomaly) embedded in a manifold of 
higher dimension $D$.

To appreciate the role of the higher-dimensional embedding, 
let us first remark that a (consistent) anomaly in $2n$ 
dimensions cannot be cancelled by adding  to the classical 
invariant action a local non-invariant $2n$-dimensional 
``counterterm'' 
$\G^{(1)}_E[A,\o]=-i \int \g[A,\o]$ that depends on the 
gauge and gravitational fields only (as does $\hat I_{2n+1}$). 
Indeed, a consistent anomaly $\hat I_{2n}^1$, characterized 
by a non-vanishing $\hat I_{2n+2}$, is only defined up to the 
addition of such a local counterterm;\,%
\footnote{\,One always has the freedom to add a local 
counterterm to the action. If this was enough to cancel 
the anomaly one could consistently quantize the theory 
without problems.}
this is the essence of the descent equations (\ref{-2piiindex}) 
or (\ref{-2piindex}). To see this, suppose one has the one-loop 
anomaly $\delta\G_E=-i\int \hat I^1_{2n}$. Upon descent this 
leads to $\hat I_{2n+2}$. If one adds the counterterm $\G^{(1)}_E[A,\o]$ 
to the classical action  the variation of the new effective action 
and the descent equations (\ref{-2piindex}) are
\ba\label{modeffaction}
\delta\G_E+\delta\G^{(1)}_E 
&=& -i\int \left( \hat I^1_{2n}+\delta\g\right),\nonumber\\
\d\left( \hat I^1_{2n}+\delta\g\right) 
&=& \delta \hat I_{2n+1} + \delta\d\g 
= \delta\left( \hat I_{2n+1} + \d\g\right),\nonumber\\
\d \left( \hat I_{2n+1} + \d\g\right) &=& \hat I_{2n+2} + 0
\ea
with the same $\hat I_{2n+2}$ as before; the invariant polynomial 
is insensitive to the addition of a local counterterm.

While addition of a local counterterm cannot eliminate the anomaly, 
it can be used to shift between two different expressions of the 
``same'' anomaly. Consider as an example the mixed U(1) 
gauge-gravitational anomaly for a negative chirality 
spin-${1\over 2}$ fermion in four Minkowskian dimensions 
characterized by the invariant 6-form
\be\label{mixed}
\hat I_6^{\rm mixed} = -{q\over 12 (4\pi)^2} \ \cF \tr R^2
\ee 
(recall that for U(1) gauge fields $A\simeq i q \cA$, $F\simeq i q\cF$\
and $v\simeq i q \tilde \e$). Upon descent, 
this gives $\hat I_4^{{\rm mixed},1}$ either as 
\be\label{mixedI41}
\hat I_4^{{\rm mixed},1}= -{q\over 12 (4\pi)^2} \, \tilde \e \tr R ^2
\quad {\rm or} \quad
\hat I_4^{{\rm mixed},1}= -{q\over 12 (4\pi)^2} \, \cF \tr \e\d\o\ .
\ee 
Addition of the counterterm
\be\label{countermixd}
\G^{(1)}= -{q\over 12 (4\pi)^2}  \int \cA \tr\! \big(\o\d\o+{2\over 3}\, \o^3\big)
\ee
allows  interpolation between the two expressions of the anomaly since\break\hfill
$\delta \G^{(1)}= -{q\over 12 (4\pi)^2}
\left( \int \cF \tr \e\d\o - \int \tilde \e\tr R^2\right)$.

The preceding discussion shows that the anomaly cannot be cancelled 
by adding local terms defined on the same $2n$-dimensional manifold 
on which live the chiral fields responsible for the anomaly. 
Instead, we will consider local terms defined on a higher-dimensional 
manifold which contains the $2n$-dimensional one as a submanifold.
 
The simplest example is a 3-dimensional manifold $M_3$ whose boundary 
is a 2-dimensional manifold $M_2=\partial M_3$. 
In practice, one has to pay attention to the orientations of 
$\partial M_3$ and $M_2$ and be careful whether what one 
calls $M_2$ is $\partial M_3$ or $-\partial M_3$, i.e.
$\partial M_3$ with opposite orientation. Suppose that on 
$M_2$ lives a chiral spin-${1\over 2}$ field coupled to a 
gauge field $A$.
The gauge anomaly is (for positive Euclidean chirality) 
$\delta\G_E = -i\int_{M_2} \hat I_2^1$ where $ \hat I_2^1$ 
is obtained by the
descent equations from
$\hat I_4= -{1\over 4\pi} \tr F^2$.
Explicitly
\ba\label{3dCS}
\hat I_3&=& -{1\over 4\pi} \tr \Big( A\d A +{2\over 3} A^3 \Big)
\equiv  -{1\over 4\pi} \,Q_3^{\rm CS}\,,\\[1mm]
\hat I_2^1&=&-{1\over 4\pi} \tr v \d A
\equiv  -{1\over 4\pi}\, Q_2^{{\rm CS},1}\label{3dCSa}
\ea
where $Q_3^{\rm CS}$ is the usual Chern--Simons 3-form, obviously obeying
\be\label{CSrel}
\delta Q_3^{\rm CS} = \d  Q_2^{{\rm CS},1} \ .
\ee
Now suppose that the 3-dimensional Euclidean action contains a 
Chern--Simons term
\be\label{CSaction}
S_{\rm CS}=-{i\over 4\pi} \int_{M_3}  Q_3^{\rm CS}\,.
\ee
As discussed in Section 2.1, this topological term needs to be 
purely imaginary in order to correspond to a real term  in the
Minkowskian action. On the other hand, being imaginary in the Euclidean 
case is exactly 
what is needed to match the anomalous part of the effective 
action, as we now proceed to show. Under a gauge variation, 
the Chern--Simons term transforms as
\be\label{CStrans}
\delta S_{\rm CS}=-{i\over 4\pi} \int_{M_3} \d  Q_2^{{\rm CS},1}\ ,
\ee
which would vanish if $M_3$ had no boundary. By Stoke's theorem we have
\be\label{CStrans2}
\delta S_{\rm CS}=-{i\over 4\pi} \int_{\partial M_3}  Q_2^{{\rm CS},1}
=i \int_{M_2} \hat I_2^1\, .
\ee
Thus the non-invariance of the Chern--Simons term is localized on 
the 2-dimensional boundary manifold $M_2$ and, with the coefficient 
chosen as above, it exactly cancels the one-loop anomaly. This is 
called anomaly cancellation by anomaly inflow from the bulk.
This example is particularly simple as the Chern--Simons term is 
nothing but $S_{\rm CS}=i\int \hat I_3$ and the anomaly inflow
is governed directly by the descent equations 
$\delta \hat I_3 = \d \hat I_2^1$.

As an example of a somewhat different type, consider a 5-dimensional 
Minkowskian theory involving a U(1)-gauge field and gravity 
and suppose it 
admits solutions that are analogous to magnetic monopoles in 4 
dimensions. In 5 dimensions these are magnetically charged 
string or vortex like solutions. Their world-volume is a 
2-dimensional manifold $W_2$. In the presence of such a 
solution, the Bianchi identity $\d\cF=0$ is modified as 
\be\label{modbian}
\d\cF = \a \ \delta^{(3)}_{W_2} \ ,
\ee
where $\a$ is some coefficient measuring the magnetic charge 
density on the string and $ \delta^{(3)}_{W_2}$ is a Dirac 
distribution 3-form  with support on the 2-dimensional 
world-volume $W_2$. It has the property that 
$ \int_{M_5} \delta^{(3)}_{W_2} \xi =  \int_{W_2} \xi$ 
for any 2-form $\xi$.  Typically, 
on $W_2$ live some chiral fields. If we suppose that they carry 
no U(1)-charge and that there are $n_+$ positive  and $n_-$ negative
(Minkowskian) chirality spin-${1\over 2}$ fields, 
there will only be a gravitational anomaly in two
dimensions equal to 
\be\label{2dgravan}
\delta\G_M = {n_- - n_+\over 96\pi} \int_{W_2} \tr \e\,\d\o \ .
\ee 
This can again be 
cancelled by anomaly inflow from the bulk. Suppose there 
is a topological term in the 5-dimensional action involving $\cF$ 
and the gravitational Chern--Simons 3-form,
\be\label{5dtop}
S_{\rm top} = \b \int_{M_5} \cF\, 
\tr\Big( \o\d\o+{2\over 3} \o^3\Big) \, .
\ee
Its variation is (using again a descent relation
 analogous to (\ref{CSrel}))
\ba\label{5dtopvar}
\delta S_{\rm top} &=& \b \int_{M_5} \cF\, \d \tr\e\,\d\o \
=\ - \b \int_{M_5} \d\cF\, \tr\e\d\o 
\nonumber\\[1mm]
&=& - \a\b \int_{M_5}  \delta^{(3)}_{W_2}  \, \tr\e\,\d\o\
=\ - \a\b \int_{W_2} \tr\e\d\o \, ,
\ea
and, if $\a\b={n_- - n_+\over 96\pi}$, this cancels the 
gravitational anomaly on the two-dimensional world-volume. 
This second example is a very simplified version of the 
cancellation of the five-brane 
anomalies in M-theory, which will be discussed (with all its 
coefficients) in some detail below.

It is worthwhile to note a generic feature of anomaly inflow in the 
previous example. Suppose we decide to rescale the U(1)-gauge 
field by some factor $\eta$ so that $\cF\to \tilde\cF=\eta\cF$. 
Then, the coefficient $\a$ in the Bianchi identity also gets 
rescaled as $\a\to\tilde\a=\eta\a$ so that it still reads 
$\d\tilde\cF=\tilde\a\, \delta^{(3)}_{W_2} $. The coefficient 
$\b$ in (\ref{5dtop}) obviously becomes $\b\to\tilde\b=\b/\eta$, 
and $\tilde\a\tilde\b=\a\b$. We see that the anomaly 
cancelling condition  
$\a\b={n_- - n_+\over 96\pi}$ {\it is invariant under any rescalings}
as it must be since the one-loop anomaly only depends on the 
integers $n_+$ and $n_-$.

It is clear from these examples that by some mechanism or another 
the variation of a $(D>2n)$-dimensional topological term in the 
classical action gives rise to a $2n$-dimensional topological term
\be\label{sclvar}
\delta S_{\rm M}^{\rm cl} = \int_{M_{\rm M}^{2n}} D_{\rm M}^{(2n)} 
\quad\Leftrightarrow\quad
\delta S_{\rm E}^{\rm cl} = - i \ \int_{M_{\rm E}^{2n}} D_{\rm E}^{(2n)}
\ee
with $D_{\rm M}^{(2n)} = D_{\rm E}^{(2n)} \equiv D^{(2n)}$
according to  (\ref{d17}). Thus the total variation of the 
$2n$-dimensional action including the one-loop anomaly is
\be
\delta \G_{\rm M} =
\int_{M_{\rm M}^{2n}} \left( \hat I^1_{2n}+ D^{(2n)} \right)
\quad \Leftrightarrow\quad 
 \delta \G_{\rm E} = - i  \int_{M_{\rm E}^{2n}}
\left( \hat I^1_{2n}+ D^{(2n)} \right)  
\ee
(where now $ \hat I^1_{2n}$ is meant to contain all the 
contributions to the one-loop
anomaly, with all the relevant signs and factors to take 
into account the different chiralities and multiplicities).
In any case, the condition for anomaly
cancellation is the same in Euclidean and Minkowski signature,
\be\label{d26} 
\hat I^1_{2n}+ D^{(2n)} = 0 \, . 
\ee

\section{Anomaly Cancellation by Inflow in Condensed Matter:
The Quantum Hall Effect\label{qhe}}
\setcounter{equation}{0}

A most important example from condensed matter is the Quantum 
Hall Effect \cite{QHE}. The relevant geometry of a Hall sample 
is two-dimensional with  a one-dimensional boundary, e.g. an 
annulus. Typically, the boundary has
 two disconnected pieces (edges) like the inner and outer 
boundary of the annulus. Adding time, the physics is on a 
$2+1$ dimensional manifold with a 1+1 dimensional boundary. 

A magnetic field $B$ is applied perpendicular to the Hall sample 
and an electric field $(E_1,E_2)$ is present along the sample (usually 
perpendicular to the edges) resulting in a voltage drop. All 
this is again described by a 2+1 dimensional electromagnetic 
field $\cA_\m$ (vector potential) with field strength 
$\cF_{\m\n}=\partial_\m\cA_\n-\partial_\n\cA_\m$ such that
(recall that our signature is $(-++)$)
\be\label{emfield}
\cF_{01}=-E_1 \ , \quad 
\cF_{02}=-E_2 \ , \quad 
\cF_{12}= B \ .
\ee
When the filling factor (controlled by the ratio of the 
electron density and the magnetic field) takes values in 
certain intervals, one observes a vanishing longitudinal 
resistivity. The conductivity matrix being the inverse of 
the resistivity matrix, the longitudinal conductivity also 
vanishes and the current and electric field are related as
\be\label{jerel}
j^a=\s^{ab} E_b = -\s^{ab} \cF_{0b} \ , \quad a,b= 1,2 
\ee
with $\s^{11}=\s^{22}=0$ and $\s^{12}=-\s^{21} \equiv \s_H$ 
being the transverse or Hall conductivity. In the integer 
Quantum Hall Effect, this Hall conductivity $\s_H$ is an 
integer multiple of $e^2/h$, or since we have set $\hbar=1$,
\be\label{condquant}
\s_H = n\ {e^2\over 2\pi} \ , \quad n\in {\bf Z} \ ,
\ee
$-e$ being the elementary charge of the electron. In the 
fractional Quantum Hall Effect, $n$ is replaced by certain 
rational numbers.

The integer Quantum Hall Effect is quite well understood 
in terms of elementary quantum mechanics of electrons in 
a strong magnetic field, giving rise to the usual Landau 
levels, together with an important role played by disorder 
(impurities) in the sample, leading to localization (see e.g. 
\cite{QHE}). The fractional Quantum Hall Effect is more 
intriguing and has given rise to a large literature (which 
I will not cite). In both cases, effective field theories 
of the Chern--Simons type have played an important role, see 
e.g. Refs. \cite{HALP1,HALP2,LOP,BAL1,BAL2,FRO}.

Here, we will only consider a simple field theoretic model 
neglecting most of the subtleties discussed in the above-mentioned 
references, as well as in others. Consider an effective field 
theory given by a $2+1$ dimensional Chern--Simons term of the 
electromagnetic vector potential $\cA_\m$ plus a 
coupling to the electromagnetic current $j^\m$,
\be\label{qhecs}
S_{2+1} = {\s\over 2} \int_{M_{2+1}} \!
\d^3 x\, \e^{\m\n\r} \cA_\m \partial_\n \cA_\r
+ \int_{M_{2+1}}\! \d^3 x\, j^\m \cA_\m \ .
\ee
(For simplicity we assume a trivial metric.) Varying this action 
with respect to $\cA_\m$ gives the equation of motion
\be\label{qheeom}
j^\m = - {\s\over 2}\,  \e^{\m\n\r} \cF_{\n\r} \ , \quad \m=0,1,2 \ .
\ee
Specializing to $\m=1,2$ and using $\e^{012}=-1$ (see Eq.\,%
(\ref{d2})) we see that the effective action (\ref{qhecs}) correctly 
reproduces the Hall relation 
(\ref{jerel}) with
\be\label{ssh}
\s_H=\s \ .
\ee
The action (\ref{qhecs}) can be rewritten using forms (cf (\ref{d3}))
as
\be\label{qhecsform}
S_{2+1} = - {\s_H\over 2} \int_{M_{2+1}} \!
 \cA\w \d \cA
+ \int_{M_{2+1}}\! {}^* j \w \cA \ .
\ee

It is well-known in the integer Quantum Hall Effect that there 
are chiral massless excitations on the boundaries (edge currents). 
They can be viewed as excitations of the incompressible 
two-dimensional electron gas or resulting from an interruption 
of the semiclassical cyclotron trajectories by the edges \cite{QHE}. 
In 
any case, they are 1+1 dimensional chiral degrees of freedom. In 1+1 
dimensions it does not matter whether they are described as chiral 
bosons or as chiral fermions, both descriptions being related. 
Suppose there are $n_k$ species of them on the edge $k$ (we label 
the two edges as $k=1,2$). Note that all species on a given edge 
have the same chirality. These chiral fermions being charged have 
a one-loop U(1) gauge anomaly. Recall that for U(1) gauge fields
we replace $A\simeq i q \cA$ and similarly for the field strength
$F\simeq i q \cF$ and for the 
gauge variation parameter $v\simeq i q \tilde\e$, where $q=-e$ is 
the (negative) electron charge. Then, 
$\tr v\,\d A \simeq -e^2 \, \tilde \e\, \d \cA$, and according 
to the general 
results of the previous section, the anomalous variation of 
the effective action on the $k^{\rm th}$ edge is
\be\label{edgeanom}
\delta_{\tilde\e} \G^{{\rm edge}\, k}
=\pm n_k \int_{M_{1+1}^{(k)}} \hat I_2^{1, spin{1\over 2}}
=\pm n_k \, {e^2\over 4\pi} \int_{M_{1+1}^k} \tilde\e\, \d\cA \ ,
\ee
where the $\pm$ accounts for the (unspecified) chirality,\footnote{\,%
One should also be careful about the orientations of $M_{2+1}$ 
and of the edges $M_{1+1}^k$ to get the signs straight.
} and we 
have used Eq.\,(\ref{3dCSa}).

On the other hand, the bulk action $S_{2+1}$ is also 
anomalous due to the boundary and it gives an anomaly inflow
\be\label{qheinflow}
\delta S_{2+1} = - {\s_H \over 2} \int_{M_{2+1}} \d ( \tilde\e\, \d\cA)
= \sum_{{\rm edges}\ k} \left( - {\s_H\over 2}\right) 
 \int_{M_{1+1}^{(k)}} \tilde\e\, \d\cA \ .
\ee
The quantum anomalies (\ref{edgeanom}) and the anomaly inflow
(\ref{qheinflow}) cancel if and only if
$\s_H = (\pm n_k)\, {e^2\over 2\pi}$. Since the anomaly 
should cancel on both edges $k$, this shows that 
$\pm n_1=\pm n_2 \equiv n$ and
\be\label{cancelqhe}
\s_H =  n\, {e^2\over 2\pi} \ , \quad n\in {\bf Z}
\ee
in agreement with Eq.\,(\ref{condquant}).
Anomaly cancellation by inflow from the bulk forces the 
Hall conductivity to be correctly quantized!

As already noted, the fractional Quantum Hall Effect is 
more complicated and the edge excitations are described by 
more complicated quasiparticles involving exotic spins and 
statistics, so that our simple argument needs to be refined. 
Somewhat related arguments can be found in \cite{BAL1,BAL2,FRO}. 
Other examples in 1+1 dimensional condensed matter where 
anomaly arguments play a role are quantum wires \cite{FRO} 
and presumably vortices, as well as defect surfaces in 
3-dimensional Euclidean statistical systems. Due to lack 
of competence, I will discuss none of them here.


\section{Examples of Anomaly Cancellation by Inflow 
in M-Theory\label{mtheory}}
\setcounter{equation}{0}

\subsection{The low-energy effective action of 
M-Theory\label{mtheoryaction}}

M-theory has emerged from a web of dualities between superstring 
theories. In its eleven-dimensional uncompactified version it can 
be considered as the strong-coupling limit of type IIA superstring 
theory. This tells us that its low-energy effective action is 
that of eleven-dimensional supergravity first written by Cremmer, 
Julia and Scherk \cite{CJS}.  In Minkowski space its bosonic part reads
(using our conventions as exposed in Section \ref{conventions})
\be\label{d10}
S_{\rm M}^{\rm CJS}={1\over 2 \k^2}\left(
\int \d^{11}x \rg\ \R
- {1\over 2}\int G\w \st G - {1\over 6} \int C\w G\w G \right)\,,
\ee
where
$\k\equiv \k_{11}$ is the 11-dimensional gravitational constant,
$\R$ is the Ricci scalar, and $G=\d C$. The coefficients of the 
second and third term in this action can be changed by rescaling 
the $C$-field.
Also, some authors use a different relation between $G$ and $\d C$. 
These issues have been extensively discussed in \cite{BM} where a 
table is given summarizing the conventions of various authors. 
Here, however, we will use the simple choice made in Eq.\,(\ref{d10}) 
which in the notation of Ref. \cite{BM} corresponds to $\a=\b=1$. 
Note that the third term is a topological term, usually referred 
to as the Chern--Simons term. 

The $C$-field equation of motion is
\be\label{d12}
\d\st G + {1\over 2}\, G\w G = 0 \ ,
\ee
or in components
\be\label{d11}
\nabla_\m G^{\m\n\r\s} + {1\over 2 \cdot 4!\cdot 4!}\
\e^{\n\r\s\m_1\ldots\m_8}\, G_{\m_1\ldots\m_4}
G_{\m_5\ldots\m_8} = 0\ ,
\ee
and the Einstein equations are
\be\label{d13}
\R_{\m\n}= {1\over 12} \Big( G_{\m\r\l\s}\, G_\n^{\ \r\l\s}
-{1\over 12}\, g_{\m\n}\, G_{\r\l\s\k}\, G^{\r\l\s\k} \Big)\,.
\ee

Just as superstring theory possesses various  D-branes,
M-theory has two fundamental branes: membranes (2-branes) 
and 5-branes.
Also, the low energy-effective action (\ref{d10}) certainly does receive
higher-order corrections. Note that in eleven-dimensional 
supergravity there is no parameter besides the gravitational 
constant $\k$, and higher order necessarily means higher order in 
$\k$. The first such term is the famous Green--Schwarz term, initially
inferred from considerations of anomaly cancellation on 
5-branes by inflow 
\cite{DLM,VW}. It reads (in Minkowski space)
\be\label{d14a}
S_{\rm GS}=- \e\ {T_2\over 2\pi} \int C\w X_8
= - \e\ {T_2\over 2\pi} \int G\w X_7
\ee
where we assumed that one can freely integrate by parts (no boundaries
or singularities), and where
\be\label{d14b}
X_8=\d X_7 = {1\over (2\pi)^3\, 4!}
\left( {1\over 8} \tr R^4 - {1\over 32} (\tr R^2)^2 \right) .
\ee
Here $T_2$ is shorthand for
\be\label{d14c}
T_2=\left( {2\pi^2\over \k^2}\right)^{1/3}
\ee
and is interpreted as the membrane tension. The parameter $\e$ can be
fixed by various considerations of  anomaly cancellation as we 
will show below.
Since there have been some ambiguities in the literature we will keep
$\e$ as a parameter and show that all anomalies considered below cancel
if and only if
\be
\e=+1\ .
\ee
Note that adding the Green--Schwarz term to the action (\ref{d10}) 
modifies the equations of motion (\ref{d12})-(\ref{d13}) by terms 
of order $\k^{4/3}$ which will be neglected below when looking for 
solutions of the ``classical'' equations of motion.

The Euclidean continuations of the action (\ref{d10}) and 
the Green--Schwarz term are
\be\label{d20} 
S_{\rm E}^{\rm CJS}=
{1\over 2 \k^2} \left( - \int \d^{11} z \, \sqrt{g}\, \R_{\rm E}
+{1\over 2} \int G_{\rm E} \w \st G_{\rm E} 
+ {i\over 6} \int C_{\rm E}\w G_{\rm E} \w G_{\rm E} \right) 
\ee
and 
\be\label{d26a} 
S^{\rm E}_{\rm GS}
= i\, \e\, {T_2\over 2\pi} \int C_{\rm E}\w X^{\rm E}_8 
=i\, \e\, {T_2\over 2\pi} \int G_{\rm E}\w X^{\rm E}_7 \, . 
\ee

\subsection{The M-Theory Five-Brane\label{5brane}}

The 5-brane and anti-5-brane are solutions of 11-dimensional
supergravity that preserve half of the 32 supersymmetries.
The metric is a warped metric preserving Poincar\'e invariance
on the $(5+1)$-dimensional world-volume (for flat 5-branes)
and the 4-form $G$ has a non-vanishing flux through any
4-sphere surrounding the world-volume. This is why the
5-branes are called ``magnetic'' sources. It will be enough
for us to exhibit the bosonic fields only.

Although the original 11-dimensional supergravity is non-chiral, 
the 5-brane is a chiral solution; it carries a chiral 
$(5+1)$-dimensional supermultiplet which gives rise to anomalies. 
Of course, the anti-5-brane carries the supermultiplet of opposite 
chirality. As a result, when computing an ``M-theory functional 
integral'' one has to sum over classical solutions of opposite 
chirality and the overall result is correctly non-chiral. However, 
we like to adopt a more modest view and consider M-theory in a given 
background with some number of 5-branes somewhere and some other 
number of anti-5-branes somewhere else. Then the anomalies 
cannot cancel between the different branes and anomaly 
cancellation must occur for each 5-brane or anti-5-brane separately.
This will be achieved by anomaly inflow from the two topological terms,
the Chern--Simons and the Green--Schwarz term.

It is not too difficult to determine the nature of the chiral 
6-dimensional supermultiplet living on the world-volume of a 
5-brane \cite{KAP}. What requires some more care is to correctly determine 
its chirality. We will see that the 5-brane acts as a ``magnetic'' 
source for the $C$-field leading to a modification of the Bianchi 
identity $\d G=0$. This is at the origin of anomaly inflow from 
the Green--Schwarz term \cite{DLM} and similar to the mechanism 
outlined in Section \ref{inflow} for the magnetic string. 
However, it was noticed \cite{W5} that there is a left-over 
``normal bundle'' anomaly which is only canceled by further 
inflow from the (slightly modified) Chern--Simons term \cite{FHMM}. 
In principal, this should have fixed the coefficient $\e$ of 
the Green--Schwarz term. In the literature one can find about 
as many times $\e=+1$ as $\e=-1$ (after eliminating the effect 
of using different conventions). This was the motivation in
\cite{BM} to redo the whole computation from first principles. 
Here we will outline this computation again, with the result $\e=+1$.

\subsubsection{The classical 5-brane solution}

We work in Minkowski space and split the coordinates into
longitudinal ones $x^\a,\ \a=0,\ldots 5$ and transverse ones
$x^m\equiv y^m,\ m=6, \ldots 10$. Then the metric is
\be\label{t1} 
\d s^2=\D(r)^{-1/3}\,
\eta_{\a\b} \d x^\a \d x^\b 
+ \D(r)^{2/3}\, \delta_{mn} \d y^m \d y^n \,,
\ee 
where 
\be\label{t1a} 
\D(r)= 1+ {r_0^3\over r^3}\,, \qquad 
 r=\left(\delta_{mn}y^m y^n\right)^{1/2}, 
\qquad   r_0\ge 0 \ , 
\ee 
(with $\eta_{\a\b}={\rm diag}(-1,1,\ldots 1)$). From this one 
has to compute the  Ricci tensor and finds that  
Einstein's equations (\ref{d13}) are solved by 
\be\label{t6}
G_{mnpq}=\pm 3\, {r_0^3\over r^5}\ \et_{mnpqs}\ y^s \, , 
\qquad {\rm all\ other\ } G_{\m\n\r\s}=0 \, . 
\ee 
The other equation of motion (\ref{d11}) reduces to 
$\partial_m\left( \rg\ G^{mnpq}\right)=0$,
which is automatically satisfied. 
The solution with the upper sign ($+$) is
called a 5-brane and the one with the lower sign ($-$) an
anti-5-brane. Details are given e.g. in \cite{BM}, where one 
can also find a discussion of how things change under a rescaling 
of the $C$-field.
The 4-form corresponding to (\ref{t6}) is
\be\label{t7} 
G=\pm\  {r_0^3\over 8 }\ \et_{mnpqs}\ 
{y^s\over r^5}\, \d y^m\w \d y^n\w \d
y^p\w \d y^q 
\ee 
and for any 4-sphere in the transverse space
surrounding the world-volume we have the ``magnetic charge''
\be\label{t7a} 
\int_{S^4} G\ =\ \pm\ 3
r_0^3 {\rm vol}(S^4) \ =\ \pm\ 8\pi^2 r_0^3 \ . 
\ee 
Hence, for the 5-brane the flux of $G$ is positive and for the
anti-5-brane it is negative.

The parameter $r_0$ sets the scale for the (anti-) 5-brane
solution. One can compute the energy per 5-volume of the brane,
i.e. the 5-brane tension $T_5$ as a function of $r_0$. Using the Dirac
quantization condition between membranes and 5-branes then relates
the membrane tension $T_2$ and the 5-brane tension $T_5$ as
$T_2\, T_5 ={2\pi\over 2\k^2}$ so that in the end
$8\pi^2 r_0^3={2\pi\over T_2}$, see \cite{BM} for details. 
(Recall from (\ref{d14c}) that
$T_2=( 2\pi^2/\k^2)^{1/3}$.)
It follows that Eq.\,(\ref{t7a}) can be rewritten as
\be\label{t10}
\int_{S^4} G\ =\ \pm\ {2\pi\over T_2} = \pm\ (4\pi \k^2)^{1/3}.
\ee
This is equivalent to the modified Bianchi identity
\be\label{t11}
\d G = \ \pm\ {2\pi\over T_2}\delta^{(5)}_{W_6}
 = \ \pm\ (4\pi \k^2)^{1/3}    \
\delta^{(5)}_{W_6}
\ee
where again the upper sign ($+$) applies for a 5-brane and 
the lower sign ($-$) for an anti-5-brane.
$\delta^{(5)}_{W_6}$ is a 5-form Dirac distribution with support
on the world-volume $W_6$ such that
$\int_{M_{11}} \o_{(6)}\w \delta^{(5)}_{W_6}
= \int_{W_6} \o_{(6)} \ .$

To summarize, the 5-brane and anti-5-brane solutions both have a
metric given by (\ref{t1}). The 4-form $G$ is given by (\ref{t7})
and satisfies the Bianchi identity (\ref{t11}). The upper sign 
always corresponds
to 5-branes and the lower sign to anti-5-branes.

\subsubsection{The zero-modes}

The (massless) fields that live on a five-brane are the zero-modes
of the equations of motion in the background of the 5-brane solution.
Hence, to determine them,
we will consider the zero-modes of the bosonic equations of motion
in this 5-brane  background. The fermionic zero-modes then are 
simply inferred from the completion of 
the supermultiplet. The anti-5-brane
background can be treated similarly (flipping signs in appropriate
places).

Apart from fluctuations describing the position of the 5-brane,
there are zero-modes of the $C$-field. A zero-mode is a
square-integrable fluctuation $\delta G=\d \delta C$ around the
5-brane solution $G_0$ (given by (\ref{t6}) or (\ref{t7}) with the
upper sign) such that $G=G_0+\delta G$ still is a solution of
(\ref{d11}) or (\ref{d12}). Of course, $G$ must also solve the
Einstein equations to first order in $\delta G$. This will be the
case with the same metric if the r.h.s. of (\ref{d13}) has no term
linear in $\delta G$.

The linearization of Eq.\,(\ref{d11}) around the 5-brane solution
(\ref{t6}) is 
\be\label{t15} 
\nabla_\m \delta G^{\m\n\r\s} +
{1\over 4! \, 4!}\, {3\, r_0^3\over r^5}\ 
\e^{\n\r\s\m_1\ldots \m_4 mnpq}\ \et_{mnpqs}\,
y^s \, \delta G_{\m_1\ldots\m_4} =0 \, . 
\ee 
Since there are only 5
transverse directions, the second term is non-vanishing only if
exactly one of the indices $\n\r\s\m_1\ldots \m_4$ is transverse.
It is not too difficult to see that the only solutions are such
that all components of $\delta G$ but $\delta G_{m\a\b\g}$ vanish.
This also ensures that $\delta G$ cannot contribute linearly to
the Einstein equations. We take the ansatz \cite{KAP}
\be\label{t16} 
\delta G_{m\a\b\g} =\D(r)^{-1-\zeta}\, r^{-5}\,
y^m\, H_{\a\b\g} \, , \quad {\rm with}\ \partial_n H_{\a\b\g}=0 \, ,
\ee 
and use $\rg=\D(r)^{2/3}$, $g^{mn}=\D(r)^{-2/3}\,
\delta^{mn}$, $g^{\a\b}=\D(r)^{1/3}\, \eta^{\a\b}$, 
as well as the convention that indices of $H_{\a\b\g}$ are raised with
$\eta^{\a\b}$  and those of $\delta G_{m\a\b\g}$ with $g^{mn}$ and
$g^{\a\b}$. This means that 
$\delta G^{m\a\b\g} = \D(r)^{-2/3-\zeta}\, r^{-5}\, y^m\, H^{\a\b\g}$. 
We further need
\be\label{t17} 
\e^{\a\b\g t\delta\e\vf mnpq}\, \et_{mnpqs} = -
{4!\over \rg}\, \delta^t_s\ \et^{\a\b\g\delta\e\vf} \, , 
\ee 
with
$\et^{\a\b\g\delta\e\vf}$ completely antisymmetric and
$\et^{012345}=-1$, i.e. $\et$ is exactly the $\e$-tensor (as
defined in (\ref{d2})) for the $(5+1)$-dimensional world-volume
with metric $\eta_{\a\b}$. Then, for $(\n,\r,\s)=(\a,\b,\g)$, Eq.\,%
(\ref{t15}) becomes\,\footnote{\,For $(\n,\r,\s)=(m,\b,\g)$ Eq.\,%
(\ref{t15}) gives $\partial_{\a}H^{\a\b\g}=0$, so that
$H_{\a\b\g}=3\ \partial_{[\a}B_{\b\g]}$, as expected.}
\be\label{t18}
\partial_m\left( \D(r)^{-\zeta}\, r^{-5}\, y^m\right) H^{\a\b\g}
- {r_0^3\over 2}\,
\et^{\a\b\g\delta\e\vf}
\D(r)^{-1-\zeta}\, r^{-8}\, H_{\delta\e\vf} = 0 \, .
\ee
Since
$\ \partial_m\left( \D(r)^{-\zeta}\, r^{-5}\, y^m\right)$
$= + 3\, \zeta\, \D(r)^{-\zeta-1}\, r_0^3\, r^{-8}$ we finally get
\be\label{t19}
\zeta\ H^{\a\b\g} = {1\over 6} \  \et^{\a\b\g\delta\e\vf}\
H_{\delta\e\vf} \, .
\ee
Consistency of this equation requires either $\zeta=+1$ in which
case $H$ is self-dual (cf (\ref{d33})) or $\zeta=-1$ in which case
$H$ is anti-self-dual.

As mentioned above, the zero-modes must be square-integrable,
\ba\label{t20} 
\infty 
&>& 
\int\! \d^{11}x \rg\, \delta
G_{m\a\b\g}\,\delta G^{m\a\b\g} \nonumber\\
&&=
{8\pi^2\over 3}\! \int_0^\infty\!\! \d r\, r^{-4}\D(r)^{-1-2\zeta} \!
\int_{W_6}\!\d^6 x\, H_{\a\b\g}\, H^{\a\b\g} \, . 
\ea
The $r$-integral converges if and only if
$\zeta>0$. Thus square-integrability selects $\zeta=+1$ and,
hence, $H=\d B$ is a self-dual 3-form on the world-volume.

To summarize, in Minkowski signature, on a 5-brane, there is a
self-dual 3-form $H$ (which continues to an anti-self-dual
Euclidean 3-form $H_{\rm E}$), while on an anti-5-brane the 3-form
$H$ is anti-self-dual (and continues to a self-dual Euclidean
3-form $H_{\rm E}$). To complete the 6-dimensional
supermultiplets, we know that  the self-dual 3-form is accompanied
by two spinors of positive chirality, and the anti-self-dual
3-form by two spinors of negative chirality. We note that the 
same discussion can be equally well carried out entirely in the 
Euclidean case (see \cite{BM}), with the same result, of course.

\subsubsection{The tangent and normal bundle anomalies}

Now that we have determined the nature and chiralities of the fields
living on the 5-brane world-volume, it is easy to determine the 
one-loop anomaly, using the results of Section \ref{oneloop}. 
For the Euclidean 5-brane we have an anti-self-dual 3-form and 
two negative chirality spinors. While
the 3-form cannot couple to gauge fields, the spinors couple to
the ``$SO(5)$-gauge" fields of the normal bundle. This coupling
occurs via
\be
D_i=\partial_i+{1\over4}\,\omega_{ab,i}\gamma^{ab}
+{1\over4}\,\omega_{pq,i}\gamma^{pq}
\ee
inherited from the eleven-dimensional spinor. Here $a,b$ and $i$
run from 1 to 6, while $p,q = 7,\ldots 11$. Thus $\omega_{pq,i}$
behaves as an $SO(5)$-gauge field $A^\a_i$ with generators
$\l^\a\sim{1\over2}\g^{pq}$. We see that the relevant $SO(5)$
representation is the spin representation \cite{W5} and hence
($R_{pq}=d\o_{pq}+\o_{pr}\o_{rq}\equiv R_{pq}^\perp$)
\begin{eqnarray}
F=F^\a\l^\a&\longleftrightarrow&{1\over4}R_{pq}^\perp\g^{pq}\\[1mm]
{\rm ch}(F)&\longleftrightarrow&
\tr\exp\left({i\over2\pi}{1\over4}R_{pq}^\perp\g^{pq}\right)
\equiv{\rm ch}(S(N))\,.
\end{eqnarray}
This trace appeared already in (\ref{indDA}), except that there
$R_{ab}$ was the curvature on the manifold (i.e. on the tangent
bundle). One has
\be
{\rm ch}(S(N))
=
4\left[1\!-\!{1\over(4\pi)^2}\,{1\over4}\tr R_\perp^2
\!+{1\over(4\pi)^4}\,\Big[\!-\!{1\over24}\tr R_\perp^4
\!+\!{1\over32}\,(\!\tr R_\perp^2)^2\Big]\!+\ldots\right].
\ee
The relevant anomaly polynomial includes an extra factor
${1\over2}$ from a chirality projector (as in (\ref{indDA}))
as well as a minus sign for negative chirality. It is
($R=\tilde R + R_\perp$)
\begin{eqnarray}
\left[-{1\over2}\ \hat A(M_6)\ {\rm ch}(S(N))\right]_8
=-{2\over(4\pi)^4}&& \hskip-4.mm
\left[ {1\over360}\tr \tilde R^4
+{1\over288}(\tr \tilde R^2)^2\right.
\nonumber\\[1mm]
&& \hskip-5.mm \left.-{1\over24}\tr R_\perp^4
+{1\over32}(\tr R_{\perp}^2)^2
-{1\over48}\tr \tilde R^2\tr R_\perp^2\right].
\end{eqnarray}
The part not involving $R_\perp$ is just $-2[\hat A(M_6)]_8$ and
can be interpreted as the contribution to the tangent bundle
anomaly of the two negative chirality spinors on $M_6$. Adding the
contribution of the anti-self-dual three-form, which is
$\left[-\left(-{1\over8}\right)L(M_6)\right]_8$
(evaluated using $\tilde R$) we get the anomaly
on the Euclidean 5-brane as $\delta \G_E=-i\int \hat
I_6^{1,5-brane}$ with
\begin{eqnarray}\label{I85bran}
\hat I_8^{5-brane}&=&2\pi\left[-{1\over2}\ \hat A(M_6)\
{\rm ch}(S(N))+{1\over8}L(M_6)\right]_8
\nonumber\\
&=&-X_8(\tilde R)-\hat I_8 ^{normal}\,,
\end{eqnarray}
where $X_{8}$ is given in (\ref{d14b}) (now with
$R\rightarrow\tilde R$) and
\begin{equation}
\hat I_8 ^{normal}
={1\over(2\pi)^34!}\left[-{1\over8}\tr R_\perp^4
+{3\over32}(\tr R_\perp^2)^2
-{1\over16}\tr \tilde R^2\tr R_\perp^2\right].
\end{equation}
The part $-X_8(\tilde R)$ is called the tangent bundle anomaly and
$-\hat I_8^{normal}$ the normal bundle anomaly.

\subsubsection{Anomaly inflow from the Green--Schwarz and 
Chern--Simons terms}

In this subsection we return to Minkowski space. As we have seen, the
5-brane has chiral zero-modes on its 6-dimensional world-volume
with its Minkowski anomaly given by
\begin{equation}\label{I6brane}
\delta\G_M^{1-loop}=\int_{W_6}\hat I_6^{1,5-brane}\,,
\end{equation}
where $\hat I_6^{1,5-brane}$ is the descent of $\hat
I_8^{5-brane}$ given in (\ref{I85bran}) and
$I_8^{5-brane}=-X_8(\tilde R)-\hat I_8 ^{normal}$. The tangent
bundle anomaly $-X_8(\tilde R)$ is cancelled \cite{DLM}
through inflow from
the Green--Schwarz term $\sim\int G\w X_7(R)$. The latter, however,
gives $X_8(R)=X_8(\tilde R + R_\perp)$, not $X_8(\tilde R)$. The
difference, as well as the normal bundle anomaly is cancelled
through inflow from the Chern--Simons term as was shown in
\cite{W5, FHMM}.  As a result, cancellation of the total 5-brane
anomaly fixes both coefficients of the Green--Schwarz and
Chern--Simons terms. In particular, it establishes a correlation
between the two coefficients.  Moreover, as we will see,
cancellation can only occur if the sign of the anomaly due to the
five-brane zero-modes is exactly as in (\ref{I6brane}),
(\ref{I85bran}).

Let us first consider the simpler inflow from the Green--Schwarz term
(\ref{d14a}) in the form $S_{GS}=-\e{T_2\over2\pi}\int G\wedge X_7$. 
Using the Bianchi identity (\ref{t11}) we get
\ba
\delta S_{GS}
&=&-\e\, {T_2\over2\pi}\int G\wedge\delta X_7\
=\ -\e\, {T_2\over2\pi}\int G\wedge\d X_6^1\nonumber\\[1mm]
&=&\e\, {T_2\over2\pi}\int \d G\wedge X_6^1\
=\ \e \int \delta^{(5)}_{W_6}\w X_6^1\
=\ \e \int_{W_6} X_6^1\ ,
\ea
where, as already noted, $X_6^1$ is $X_6^1(R)$.
This corresponds via
descent to an invariant polynomial
\begin{equation}
\hat I_8^{GS}=\e \ X_8(R).
\end{equation}

Next, inflow from the Chern--Simons term is more subtle. We review
the computation of \cite{FHMM}, again paying particular attention
to issues of signs and orientation. The two key points in
\cite{FHMM} are: (i) the regularization
\begin{equation}
\delta^{(5)}_{W_6}\rightarrow \d\rho\wedge{e_4\over2}
\end{equation}
where $\rho(r)$ rises monotonically from $-1$ at $r=0$ to 0 at
some finite distance $\tilde{r}$ from the 5-brane, and $e_4=\d
e_3$ is a certain angular form with $\int_{S^4} {e_4\over2}=1$;
and (ii) a modification of the Chern--Simons term close to the
5-brane, where $G\neq\d C$.

The regularized Bianchi identity reads
\begin{equation}
\d G={2\pi\over T_2}\,\d \rho \wedge
{e_4\over2}
\end{equation}
which is solved by (requiring regularity at $r=0$ where $e_4$ is
singular)
\begin{eqnarray}
G&=&\d C + {\pi\over T_2}
(2\d\rho\wedge\d B-\d\rho\wedge e_3)
\nonumber\\
&=&{\pi\over T_2}\rho\, e_4
+\d\left( C-{\pi\over T_2}
(\rho\, e_3+2\d\rho\w B) \right)
\nonumber\\[1mm]
&\equiv& {\pi\over T_2}\rho\, e_4
+\d\,\Ct \ .
\end{eqnarray}
Under a local Lorentz transformation, $\delta e_3=\d e_2^1$, and
$G$ is invariant if $\delta C=0$ and $\delta B={1\over 2}\,e_2^1$.
Note that \cite{FHMM} include the $\d\rho\wedge B$-term in $C$ and
hence get a non-trivial transformation for $C$. If we let
$\Gt=\d\Ct$ then the modified Chern--Simons term is
\begin{equation}\label{SFHMM}
\St_{CS}=-{1\over12\k^2}\lim_{\e\to 0}
\int_{M_{11}\backslash D_{\e}W_6}\Ct\w\Gt\w\Gt\,,
\end{equation}
where $M_{11}\backslash D_{\e}W_6$ is $M_{11}$ with a small
``tubular" region of radius $\e$ around the 5-brane world-volume
cut out. (Of course, this radius $\e$ should not be confused with
the $\e$ which is the coefficient of the Green--Schwarz term.)
Its boundary is
\begin{equation}\label{boundary}
\partial(M_{11}\backslash
D_{\e}W_6)=-S_{\e}W_6
\end{equation}
where $S_{\e}W_6$ is the 4-sphere bundle over $W_6$. Note the
minus sign that appears since the orientation of the boundary is
opposite to that of the sphere bundle.

Under a local Lorentz transformation $G$ and hence $\Gt$ are
invariant and
\begin{equation}\label{VarCt}
\delta \Ct =-{\pi\over T_2}\,
\d (\rho\, e_2^1)\,.
\end{equation}
Inserting this variation into (\ref{SFHMM}), and using $\d \Gt=0$
one picks up a boundary contribution\footnote{\,%
We get three minus signs, one from (\ref{SFHMM}),
(\ref{boundary}) and (\ref{VarCt}) each.
Apparently the one from (\ref{boundary}) was overlooked in
\cite{FHMM}.
}
\begin{equation}
\delta\St_{CS}=-{\pi\over 12\k^2\,T_2}\ \lim_{\e\to 0}\, 
\int_{S_{\e}W_6}\rho e_2^1\w\Gt\w\Gt\,.
\end{equation}
In $\Gt=\d C-{\pi\over T_2} (\d\rho\w
e_3+\rho\, e_4-2\d \rho\w\d B)$ the terms $\sim\d\rho$ cannot
contribute to an integral over $S_{\e}W_6$. Also the contribution
of the $\d C$-terms vanishes in the limit $\e\to 0$. Hence the
only contribution comes from \cite{FHMM,BOTT}
\begin{equation}
\int_{S_\e W_6} e_2^1\w e_4\w e_4=2 \int_{W_6} p_2(NW_6)^1\,,
\end{equation}
where $p_2(NW_6)^1$ is related  via descent to the second
Pontrjagin class $p_2(NW_6)$ of the normal bundle given below.
Using $\rho(0)=-1$ and (\ref{d14c}) we arrive at
\begin{equation}
\delta\St_{CS}={1\over6\k^2}
\left({\pi\over T_2}\right)^3 \int_{W_6} p_2(NW_6)^1
={\pi\over 12}
\int_{W_6} p_2(NW_6)^1 \, .
\end{equation}
This corresponds to an invariant polynomial
\begin{equation}
\hat I_8^{CS}={\pi\over12}\,
p_2(NW_6)\, .
\end{equation}
Using
\begin{eqnarray}
{\pi\over12}\,p_2(NW_6) 
&=&
{1\over(2\pi)^34!}
\left(-{1\over4}\tr R_{\perp}^4
+{1\over8}(\tr R_{\perp}^2)^2\right)
\nonumber\\[1mm]
X_8(R)
&=&
X_8(\Rt)+{1\over(2\pi)^34!}
\left({1\over8}\tr R_{\perp}^4-{1\over32}(\tr R_{\perp}^2)^2
-{1\over16}\tr \Rt^2\tr R_{\perp}^2\right)
\end{eqnarray}
we find that the total inflow corresponds to
\be
\hat I_8^{GS}+\hat I_8^{CS}
=\e\,X_8(\Rt)
+{1\over(2\pi)^3 4!}
\bigg[\left({\e\over8}-{1\over 4}\right)\tr R_{\perp}^4
+\left({1\over 8}-{\e\over32}\right)(\tr R_{\perp}^2)^2
-{\e\over16}\tr \Rt^2\tr R_{\perp}^2\bigg].
\label{totalinflow}
\ee

Now it is easy to study anomaly cancellation. Invariance of the
full quantum effective action requires that the sum of
(\ref{I85bran}) and (\ref{totalinflow}) vanishes. This gives four
equations
\ba
\hskip6.mm
\e=1\,,\quad  &&\quad
\left({\e\over8}-{1\over 4}\right)
=-{1\over8}\,,\nonumber\\
\left({1\over8}
-{\e\over32}\right)
={3\over32}\,,
\quad &&\quad 
 -{\e\over16}=-{1\over16}\ .
\label{cancellationcond}
\ea
The first equation ensures the
cancellation of the tangent bundle anomaly and 
the three other equations ensure the
cancellation of the normal bundle anomaly. All four equations
are solved by
\begin{equation}\label{GSsign}
\e= +1\, .
\end{equation}
It is quite amazing to see that anomaly cancellation requires 
four different terms to vanish, and they all do if the single 
coefficient $\e$ is chosen as above. Note also that a rescaling 
of the $C$-field changes the coefficients of the Chern--Simons 
and Green--Schwarz terms, but cannot change the {\it relative} 
sign between them. The effect of such rescalings has been 
carefully traced through the computations in ref \cite{BM} where
it can be seen that the resulting equations (\ref{cancellationcond})
are indeed invariant under these rescalings, as they should.

It is also interesting to
note that the four conditions (\ref{cancellationcond}) for anomaly
cancellation have enough structure to provide a check that we
correctly computed the sign of the one-loop anomaly (if we believe that the anomaly must cancel). Suppose we
replaced equation (\ref{I85bran}) by
\begin{equation}
\hat I^{5-brane}_8(\eta)=-\eta\,[X_8(\Rt)+\hat I_8^{normal}]\, ,
\qquad \eta=\pm 1\,.
\end{equation}
Then equations (\ref{cancellationcond}) would get an extra factor
$\eta=\pm 1$ on their right-hand sides.
However, the four equations are enough to uniquely determine both 
$\e=+1$ and $\eta=+1$. Said differently, a one-loop anomaly of
opposite sign could not be cancelled through inflow from the
Chern--Simons or Green--Schwarz terms even with their signs flipped.
At first sight this might seem surprising. However, as
we have seen, such a sign flip merely corresponds to a
redefinition of the fields and obviously cannot yield a different
inflow.

\subsection{M\,-Theory on $S^1/Z_2$: the Strongly-Coupled 
Heterotic String\label{s1z2}}

While compactification of M-theory on a circle $S^1$ leads to 
(strongly-coupled) type IIA superstring theory, compactification 
on an interval gives the strongly-coupled heterotic string \cite{HW}. 
There are two ways to view this latter compactification. 
On the one hand,
one considers the compactification manifold as being 
ten-dimensional Minkowski space $M_{10}$ times the interval 
so that the 11-dimensional space-time has two boundaries,
each of which is a copy of  $M_{10}$. This is called the 
``downstairs approach''. On the other hand, the interval 
being $S^1/{\bf Z}_2$, one may start with the 11-dimensional 
manifold being $M_{10}\times S^1$ and then perform the ${\bf Z}_2$ 
orbifold projection. In this case there are no boundaries, 
but two orbifold fixed-planes, each of which is again a copy 
of $M_{10}$. This is called the ``upstairs approach''.

One may also consider more complicated compactifications
on orbifolds like e.g. $T^5/{\bf Z}_2$ with many intersecting 
orbifold planes. The latter constructions have given rise to 
some model building,
see e.g. \cite{Ovrut}.

Here we will work in the upstairs approach. As argued in \cite{HW}
the orbifold projection eliminated half of the supersymmetry 
leaving only one chiral (ten-dimensional) gravitino on each 
of the ten-dimensional orbifold planes. This leads to a 
gravitational anomaly with an irreducible $R^6$ piece. 
The latter piece can be cancelled by adding $E_8$ gauge 
fields on each of the orbifold planes (interpreted as 
``twisted'' matter). The total one-loop anomaly then no 
longer has this $R^6$ piece and, remarkably, has a factorized 
form on each of the planes, a necessary condition for anomaly 
cancellation by inflow from the Green--Schwarz and Chern--Simons terms.
There has been a long series of papers discussing this cancellation
that culminated with Ref. \cite{BDS}, each paper correcting some 
errors of the preceding ones. However, this was not the end 
of the story, since one of the authors of  \cite{BDS} realized 
that there was still an unnoticed numerical error, and to 
correctly obtain complete anomaly cancellation requires a 
slight modification of the Chern--Simons term in the vicinity of the 
orbifold planes, quite similar 
to what happened for the 5-brane as discussed above. 
This was reported in \cite{BM} and we will review these 
results in this subsection. The attitude taken in \cite{BM} was to
show that anomaly cancellation in this case again determines 
the value of $\e$ to be $+1$. Here, instead, we will consider 
that the coefficient of the Green--Schwarz-term is already fixed 
from the 5-brane anomaly cancellation and that with this value 
we correctly obtain anomaly cancellation also in the present 
case.

\subsubsection{The one-loop anomalies on the orbifold 10-planes}

As always in Minkowski signature, we label the coordinates
as $x^\m, \ \m=0,\ldots 10$. Here we will distinguish the circle 
coordinate $x^{10}\in[-\pi r_0,\pi r_0]$ from the other 
$x^{\bar{\m}} ,\ \bar\m=0,\ldots 9$.
The ${\bf Z}_2$-projection then acts as $x^{10}\to - x^{10}$.
As one can see from the Chern--Simons term,
$C_{\bar{\m}\bar{\n}\bar{\r}}$ is ${\bf Z}_2$-odd and
$C_{\bar{\m}\bar{\n} 10}$ is ${\bf Z}_2$-even
($\bar\m,\bar\n,\bar{\r}=0,\ldots 9$). The projection on
${\bf Z}_2$-even fields then implies e.g. that
\begin{equation}
C=\tilde B \w\d x^{10}\ ,
\end{equation}
and all other components of $C$ projected out. Also, this
${\bf Z}_2$-projection only leaves half of the components of the
eleven-dimensional gravitino \cite{HW}. What remains is a
ten-dimensional gravitino of positive chirality (in
Minkowskian space), together with one negative chirality
spin-${1\over2}$ field. Of course,  in Euclidean space, this
corresponds to one negative chirality spin-$3\over2$ and a
positive chirality spin-$1\over2$ fermion. The 1-loop anomaly due
to the eleven-dimensional gravitino on each 10-plane $M_{10}^A$,
$A=1,2$ is thus given by
\begin{equation}
\hat I_{12,A}^{gravitino}={1\over2}\cdot{1\over2}\left(-\hat
I_{12}^{spin{3\over2}}(R_A)+I_{12}^{spin{1\over2}}(R_A)\right)\ ,
\end{equation}
where one factor ${1\over2}$ is due to the Majorana condition and
the other factor $1\over2$ due to the ``splitting" of the anomaly
between the two fixed planes \cite{HW}. $R_A$ denotes the
curvature two-form on $M_{10}^A$ which simply is the
eleven-dimensional curvature $R$ with its components tangent to
$S^1$ suppressed. As is well known, such a polynomial has a $\tr
R^6$-piece, and one must add an $E_8$ vector multiplet in the
adjoint representation ($\Tr {\bf 1}=248$) with positive chirality
(Minkowskian) Majorana spinors on each 10-plane. Then on
each plane $M_{10}^A$ one has a 1-loop anomaly corresponding to
\begin{eqnarray}
\hat I_{12,A}&=&{1\over4}
\left(-\hat I_{12}^{spin{3\over2}}(R_A)
+I_{12}^{spin{1\over2}}(R_A)\right)
-{1\over2}\hat I_{12}^{spin{1\over2}}(R_A, F_A)
\nonumber\\[1mm]
&=&I_{4,A}\left[X_8(R_A)+{\pi\over3}\,I_{4,A}^2\right] ,
\label{factoran}
\end{eqnarray}
where we used $\Tr F_A^4={1\over100}(\Tr F_A^2)^2$,
$\Tr F_A^6={1\over7200}(\Tr F_A^2)^3$ and defined\,\footnote{\,%
$I_{4,A}$ is
exactly what was called $\tilde I_{4,i}$ in \cite{BDS}.
}
\begin{equation}
I_{4,A}={1\over(4\pi)^2}\left({1\over30}\Tr F_A^2
-{1\over2}\tr R_A^2\right)
\equiv{1\over(4\pi)^2}\left(\tr F_A^2-{1\over2}\tr R_A^2\right) .
\end{equation}
Note that in the small radius limit with $R_1=R_2=R$ one has
\ba\label{smallrad}
\left[\hat I_{12,1}+\hat I_{12,2}\right]\Big\vert_{R_1=R_2=R}
&=&\left(I_{4,1}+I_{4,2}\right)
\left[ X_8(R) +{\pi\over 3}
\left(I_{4,1}^2+I_{4,2}^2-I_{4,1}I_{4,2}\right) \right]
\nonumber\\[1mm]
&\equiv&\left(I_{4,1}+I_{4,2}\right)\, \widehat X_8(R,F_1,F_2)\, ,
\ea
thanks to the algebraic identity $a^3+b^3=(a+b)(a^2+b^2-ab)$.
Here $\widehat X_8$ is the relevant 8-form that appears
in the anomaly-cancelling term of the heterotic string,
\ba\label{hetX8}
\widehat X_8(R,F_1,F_2) = {1\over (2\pi)^3 4!}
&& \hskip-4.mm \left(\ {1\over 8} \tr R^4
+ {1\over 32} (\tr R^2)^2
-{1\over 8} \tr R^2 (\tr F_1^2 + \tr F_2^2) \right.
\nonumber\\[1mm]
&&\hskip-4.mm \left.+\,{1\over 4} (\tr F_1^2)^2
+{1\over 4} (\tr F_2^2)^2
- {1\over 4} \tr F_1^2 \tr F_2^2 \right) .
\ea

\subsubsection{Anomaly inflow and anomaly cancellation}

To begin with, there is a slight subtlety concerning the coefficients of
the Chern--Simons and Green--Schwarz terms in the upstairs formalism. 
To see this, we start in the downstairs formalism where $S_{CS}$ 
and $S_{GS}$ are given by integrals over an honest manifold with 
boundary which is $M_{10}$ times the interval $I=S^1/{\bf Z}_2$. 
Then clearly the coefficients must be those given in the 
preceding subsections,
\begin{equation}
S_{CS}=-{1\over12\k^2}\int_{M_{10}\times I}C\w G\w G\,, \qquad
S_{GS}=-{1\over(4\pi\k^2)^{1/3}}
\int_{M_{10}\times I}G\w X_7\, .
\end{equation}
Here $\k$ is the eleven-dimensional $\k$ as before. This can be
rewritten in the upstairs formalism by replacing
$\int_{I}\ldots={1\over2}\int_{S^1}\ldots$ and appropriately
identifying the fields so that the integrand is ${\bf Z}_2$-even.
This introduces an extra ${1\over2}$ in the coefficients. It is nevertheless customary to absorb this  ${1\over2}$ in a redefinition of $\k$ as
\begin{equation}
\k_{\rm U}^2=2\k^2\equiv 2\k_{\rm D}^2 \, .
\end{equation}
Then one has
\ba
S_{CS}&=&-{1\over12\k_{\rm U}^2}
\int_{M_{10}\times S^1}C\w G\w G\,,
\nonumber\\[1mm]
S_{GS}&=&-{1\over 2^{2/3}(4\pi\k_{\rm U}^2)^{1/3}}
\int_{M_{10}\times S^1}G\w X_7\, ,
\label{GSterm}
\ea
and the Chern--Simons term looks conventionally normalized. However,
due to the different dependence on $\k$, the Green--Schwarz term, 
when written in the upstairs formalism, has an extra factor of
$2^{-{2/3}}$. This will be important later on.

The factorized form (\ref{factoran}) of the anomaly
on each ten-plane is a necessary condition to allow
for local cancellation through inflow. Clearly, the
$I_{4,A}X_8$-term has the right form to be cancelled through inflow
from the Green--Schwarz term, provided $G$ satisfies a modified
Bianchi identity $\d G\sim\sum_{A=1,2}\delta_A\wedge I_{4,A}$,
where $\delta_A$ is a one-form Dirac distribution such that\break\hfill
$\int_{M_{10}\times S^1}\xi_{(10)}\wedge\delta_A\ 
=\int_{M_{10}^A}\xi_{(10)}$ for any 10-form $\xi_{(10)}$. 
This is equivalent to
prescribing a boundary value for $G$ on the boundary planes in the
down-stairs approach. Such a modified Bianchi identity is indeed necessary
to maintain supersymmetry in the coupled 11-dimensional
supergravity/10-dimensional super-Yang--Mills system \cite{HW}.
In principle, this allows us to deduce the coefficient
$-\zeta$ on the
right-hand side of the Bianchi identity in the upstairs approach.
It is given by $-(4\pi)^2 {\k_{\rm U}^2\over \l^2}$ where
$\l$ is the (unknown) Yang--Mills coupling constant.

Hence, we start with a Bianchi identity \cite{HW}
\begin{equation}\label{BI}
\d G=-\zeta \sum_{A=1,2}\delta_A\w
I_{4,A}\,.
\end{equation}
The variation of the Green--Schwarz term then is 
(recall $\delta X_7=\d X_6^1$)
\ba
\delta S_{GS}&=&-\ {1\over 2^{2/3}(4\pi\k^2_{\rm U})^{1/3}}
\int_{M_{10}\times S^1}G\w\d X_6^1\nonumber\\[1mm]
&=&-\ {\zeta\over 2^{2/3}(4\pi\k^2_{\rm U})^{1/3}}
\sum_A\int_{M_{10}^A} I_{4,A}\w X_6^1 \, .
\ea
Provided
\be\label{zetavalue}
\zeta=2^{2/3}(4\pi\k^2_{\rm U})^{1/3} \, ,
\ee
$\delta S_{GS}$ corresponds to an invariant polynomial
\begin{equation}
\hat I_{12}^{GS}=-\sum_{A}I_{4,A}\w X_8(R_A) \, .
\end{equation}
As promised, this cancels the part of the anomaly
(\ref{factoran}) involving $X_8$. Moreover, this cancellation is 
local, i.e. cancellation occurs on each plane separately.
We see that anomaly cancellation
fixes the value of $\zeta$ to be (\ref{zetavalue}), thereby
determining the value of the 10-dimensional Yang--Mills coupling
$\l$ in terms of the 11-dimensional gravitational coupling
$\k$. Although this latter aspect has drawn some attention,
one has to realize that the more interesting relation between
$\l$ and the 10-dimensional $\k_{10}$ involves the (unknown)
radius $r_0$ of the circle, similarly to the relation between the
type IIA string coupling constant and $\k$.

To study anomaly inflow from the Chern--Simons term we have to
solve the Bianchi identity for $G$ (as we did for the 5-brane).
This involves several subtleties, discussed at length in
\cite{BDS}. One important point was to respect periodicity in the
circle coordinate $x^{10}\in[-\pi r_0,\pi r_0]$ which led to the
introduction of two periodic ${\bf Z}_2$-odd ``step''
functions $\e_A(x^{10})$ such that
$\e_1(x^{10})={\rm sgn}(x^{10})-{x^{10}\over\pi r_0}$ and
$\e_2(x^{10})=\e_1(x^{10}\pm \pi r_0)$. They satisfy
\begin{equation}
{1\over2}\,\d\e_A=\delta_A-{\d x^{10}\over2\pi r_0}\ .
\end{equation}
Regularizing $\e_A$ (and hence $\delta_A$) properly gives
\begin{equation}\label{eed}
\delta_A\e_B\e_C\simeq{1\over3}\, \delta_A\
\delta_{BA}\delta_{CA} \, ,
\end{equation}
where $\delta_{BA}$ and $\delta_{CA}$ denote the Kronecker symbol.
When solving the Bianchi identity (\ref{BI}) one can (locally)
trade terms ${1\over2}\e_AI_{4,A}$ for terms $-\left(\delta_A-{\d
x^{10}\over2\pi r_0}\right)\o_{3,A}$, where
\begin{equation}
\d\o_{3,A}=I_{4,A}\,, \qquad \delta\omega_{3,A}=\d\o_{2,A}^1 \, ,
\end{equation}
since their difference is a total derivative ($\o_{3,A}$ is given
in terms of the Chern--Simons forms on $M_{10}^A$ and has no $\d
x^{10}$ component). This introduces an arbitrary real parameter
$b$ into the solution,
\begin{eqnarray}\label{G}
G&=&\d C-b\,{\zeta\over2}\,
\sum_A\left(\e_AI_{4,A}+ \o_{3,A}\w{\d x^{10}\over\pi r_0}\right)
+(1-b) \, \zeta\, \sum_A \delta_A \w \o_{3,A}
\nonumber\\
&=&\d\Big(C-b\,{\zeta\over2}\,\sum_A\e_A\o_{3,A}\Big)
+\zeta\,\sum_A\delta_A\w\o_{3,A}
\nonumber\\
&\equiv&\d\,\Ct+\zeta\sum_A\delta_A\w\o_{3,A} \, .
\end{eqnarray}
Since $G$ appears in the kinetic term $\sim\int G\w\st G$,
as well as in the energy-momentum tensor, it must be gauge
and local Lorentz invariant, $\delta G=0$. This is achieved if
\cite{BDS}
\begin{eqnarray}
\delta C&=&b\,\zeta\, \sum_{A=1,2}\o_{2,A}^1 \w
{\d x^{10}\over2\pi r_0}
+(1-b) \, \zeta \sum_A \delta_A \w \o_{2,A}^1
\nonumber\\
\Leftrightarrow\ \
\delta\Ct&=&\d\bigg(-b\,{\zeta\over2}\sum_A\e_A\o_{2,A}^1\bigg)
+\zeta\sum_A \delta_A\w\o_{2,A}^1\, .
\label{deltaC}
\end{eqnarray}

In \cite{BDS} several arguments were given in favor of one
particular value of $b$, namely $b=1$, since only then $G$ is
globally well-defined. Furthermore, the higher Fourier modes of
$C_{\bar\m\bar\n 10}$ are gauge invariant only for this value of
$b$, which is a necessary condition for a safe truncation to the
perturbative heterotic string. Last, but not least, it is only for
$b=1$ that $G$ has no terms involving $\delta_A$ which would lead
to divergent pieces in the kinetic term $\int G\w\st G$.
Nevertheless, we will keep this parameter $b$ for the time being
and show in the end that anomaly cancellation also requires $b=1$.

Note that, although $G\neq\d C$, we still have  $G=\d\Ct$
as long as
we stay away from the fixed planes. This
motivates us to introduce a modified Chern--Simons term similar to
what was done in Section 5 for the 5-brane or in \cite{BM}
when discussing M-theory on singular $G_2$-manifolds. We take
\begin{equation}\label{SCStilde}
\St_{CS}=-{1\over12\k^2_{\rm U}}\int_{M_{10}\times S^1}\Ct\w G\w
G \, ,
\end{equation}
which away from the fixed planes is just
$\sim\int \Ct\w\d\Ct\w\d\Ct$.
Then
\begin{eqnarray}\label{deltaCS}
\delta\St_{CS}&=&-{1\over12\k_{\rm U}^2}
\int_{M_{10}\times S^1}\delta \Ct\w G\w G
\nonumber\\
&=&-{1\over12\k^2_{\rm U}}\int_{M_{10} \times S^1}
\left[\d\bigg(-b\,{\zeta\over2}\sum_A\e_A\omega_{2,A}^1\big)\w
2\,\d\Ct\w\zeta\sum_C\delta_C\w\o_{3,C} \right.
\nonumber\\
&&\left. \hskip28.mm
+\zeta\sum_A\delta_A\w\o_{2,A}^1\w\d \Ct\w\d\Ct\right]  .
\end{eqnarray}
Note that we can freely integrate by parts (we assume that
$M_{10}$ has no boundary). Furthermore, since both $\delta_A$ and
$\d C=\d \tilde B\w\d x^{10}$ always contain a $\d x^{10}$, on the
r.h.s of Eq.\,(\ref{deltaCS}) one can replace $\d\Ct\rightarrow -
b\,{\zeta\over2}\sum_B \e_B I_{4,B}$, so that
\begin{equation}
\delta\St_{CS} =-{1\over12\k^2_{\rm U}}\, b^2 \left({\zeta^3\over
4}\right) \int_{M_{10} \times S^1}\!\sum_{A,B,C}
\big(2\e_A\e_B\delta_C+\delta_A\e_B\e_C\big)\o_{2,A}^1\w I_{4,B}\w I_{4,C}
\, .
\end{equation}
The modified Chern--Simons term contributes three terms
$\e\, \e\, \delta$. This factor of 3 was absent in \cite{BDS} where
inflow from the unmodified Chern--Simons term was computed.
Also the result of \cite{BDS} was obtained only after using
$\int_{S^1}\d x^{10}\e_A\e_B=\pi r_0(\delta_{AB}-{1\over3})$
which somewhat
obscured the local character of anomaly cancellation. Now,
however, due to the explicit $\delta_A$ one-forms, the inflow from
$\St_{CS}$ is localized on the 10-planes $M_{10}^A$. Using
(\ref{eed}) we find
\begin{equation}\label{deltaSt}
\delta \St_{CS}
=-{\zeta^3\over48\k^2_{\rm U}}\, b^2
\sum_{A=1,2}\int_{M_{10}^A}\o_{2,A}^1\w I_{4,A}\w I_{4,A} \, .
\end{equation}
Upon inserting the value of $\zeta$, equation (\ref{zetavalue}),
we see that this corresponds to an invariant polynomial
\begin{equation}
\hat I_{12}^{\widetilde{CS}}
=- b^2\, {\pi\over3} \sum_{A=1,2}I_{4,A}^3 \, .
\end{equation}
This cancels the remaining piece of the anomaly (\ref{factoran})
precisely if 
\be\label{bequ} 
b^2=1 \, . 
\ee 
As already mentioned there are many other arguments in favor of $b=1$,
but now we can conclude that also anomaly cancellation on
$S^1/{\bf Z}_2$ {\it requires} $b=1$, as argued in \cite{BDS}.\,%
\footnote{\,In \cite{BDS} inflow from the unmodified Chern--Simons term was
computed. This is three times smaller than (\ref{deltaSt}). Also
the factor $2^{2/3}$ in $\zeta$ was missing, so that the overall
inflow $\delta S_{CS}$ appeared 12 times smaller. This discrepancy
remained unnoticed since the anomaly cancellation condition was
expressed as ${(4\pi)^5\k^4b^2\over12\l^6}=1$. It is only after
relating ${\lambda^2\over\k^2}$ to the coefficient of the
Green--Schwarz term that one can use ${(4\pi)^5\k^4\over\l^6}=1$
and then ${b^2\over12}=1$ clearly is in conflict with $b=1$.\\[-3mm]~}

Thus we have shown that all the anomalies are cancelled {\it
locally} through inflow from the Green--Schwarz\,%
\footnote{\,It is
interesting to note that $\St_{GS}=-{1\over2^{2/3}(4\pi\k^2_{\rm
U})^{1/3}}\int_{M_{10}\times S^1}\Ct\w X_8$ would have led to the
same result.} 
and (modified) Chern--Simons terms with exactly the
same coefficients as already selected from cancellation of the
5-brane anomalies.

\subsubsection{Small radius limit and the heterotic anomaly cancelling term}

Finally, it is easy to show that in the small radius limit
($r_0\to 0$) the sum $S_{GS}+\St_{CS}$ exactly reproduces the
heterotic Green--Schwarz term. In this limit $X_8(R)$ and $X_7(R)$
are independent of $x^{10}$ and have no $\d x^{10}$
components. From $C=\tilde B\w\d x^{10}$ and $\delta C$ given in
(\ref{deltaC}) we identify the correctly normalized
heterotic $B$-field as the zero
mode of $\tilde B$ times ${(4\pi)^2\over\zeta}2\pi r_0\,$,
\begin{equation}\label{BdeltaB}
B={(4\pi)^2\over\zeta}\int_{S^1}\tilde B\w\d x^{10}\,,  \qquad
\delta B=(4\pi)^2\sum_A\o_{2,A}^1=\o_{2,YM}^1-\o_{2,L}^1~~
\end{equation}
where $\o_{2,YM}^1$ and $\o_{2,L}^1$ are related to $\tr F_1^2+\tr
F_2^2$ and $\tr R^2$ via descent.  Next,
using (\ref{G}) and (\ref{zetavalue}), the Green--Schwarz
term (\ref{GSterm}) gives in the small radius limit
\begin{eqnarray}
S_{GS}&\rightarrow& {1\over(4\pi)^2}\int_{M_{10}}\left(\d
B-\omega_{3,YM}+\o_{3,L}\right)\w X_7\nonumber\\[1mm]
&=&-{1\over(4\pi)^2}\int_{M_{10}}B\w X_8
-{1\over(4\pi)^2}\int_{M_{10}}(\o_{3,YM}-\o_{3,L})\w X_7\, .\ \ 
\label{SGShet}
\end{eqnarray}
The second term is an irrelevant local counterterm; its gauge and
local Lorentz variation corresponds to a vanishing $I_{12}\,$.
Such terms can always be added and subtracted. The modified
Chern--Simons term (\ref{SCStilde}) gives (using (\ref{G})
with $b=1$, (\ref{zetavalue}), (\ref{BdeltaB}) and
integrating by parts on $M_{10}$)
\be
\St_{CS}\rightarrow  - \sum_{A,B}
\int_{M_{10}}\left({\pi\over (4\pi)^2} B\w I_{4,A}\w I_{4,B}
-{2\pi\over3} \o_{3,A}\w I_{4,B}\w \sum_C\o_{3,C}\right)
\int_{S^1}\e_A\, \e_B\, {\d x^{10}\over 2\pi\, r_0} \ .
\ee
Using the relation
\begin{equation}
\int_{S^1}\e_A\ \e_B\ {\d x^{10}\over 2\pi\, r_0}
={1\over 2} \left(\delta_{AB}-{1\over3}\right)
\end{equation}
we get
\ba
\St_{CS}&\rightarrow&
-{1\over(4\pi)^2}\int_{M_{10}}B\w{\pi\over3}
\left(I_{4,1}^2+I_{4,2}^2-I_{4,1}I_{4,2}\right)
\nonumber\\[1mm]
&&\hskip-0.mm
-{2\pi\over9}\int_{M_{10}}(\o_{3,1}+\o_{3,2})
\left(\o_{3,1}I_{4,1}+\o_{3,2}I_{4,2}
-{1\over2}\o_{3,1}I_{4,2}-{1\over2}\o_{3,2}I_{4,1}\right)\,. 
\label{SCShet}
\ea
Again, the second term is an irrelevant counterterm. Summing
(\ref{SGShet}) and (\ref{SCShet}) we arrive at
(cf. (\ref{smallrad}))
\begin{equation}\label{Shet}
S_{GS}+\St_{CS}\rightarrow
S_{het}=-{1\over(4\pi)^2}\int_{M_{10}}B\w \hat X_8(R, F_1, F_2)+\
\mbox{local counterterms} \, ,
\end{equation}
where $\hat X_8(R, F_1,F_2)$ is the standard heterotic 8-form
given in (\ref{hetX8}). Equation (\ref{Shet}) is the correctly
normalized
heterotic anomaly-cancelling term.\footnote{\,In order to facilitate
comparison with \cite{GSW} we note that $\hat
X_8={1\over(2\pi)^34!}X^{GSW}_8$, and $S_{het}$ as given in
(\ref{Shet}) exactly equals minus the expression given in
\cite{GSW}. The missing minus sign in \cite{GSW} is due to a sign
error related to the subtle issues of orientation, and is
corrected e.g. when using the anomaly polynomials as given in
\cite{FLO}.}


\section{Concluding remarks: Brane World Cosmologies, etc\label{bwc}}
\setcounter{equation}{0}

We have studied anomaly cancellation by inflow from the bulk in 
two very different settings: the low-dimensional example of the 
Quantum Hall Effect and the high-dimensional examples of M-theory.
There are certainly many other examples one could cite and study.
One particularly interesting case are brane-world cosmologies. 
Here one has a 4-dimensional Minkowski manifold that is a ``brane'' 
embedded in a higher-dimensional manifold. Usually it is considered 
that the standard-model fields only live on the brane and only 
gravity propagates in the bulk. More sophisticated versions based 
on supergravity will also have certain gauge fields in the bulk 
and one can then study in the same way inflow of gauge and 
gauge-gravitational anomalies into the brane. This is somewhat 
reminiscent of what happens in the AdS/CFT correspondence where 
the five-dimensional $AdS_5$ supergravity has $SU(4)$ gauge fields  
and its action precisely involves a Chern--Simons term. On the 
boundary of  $AdS_5$ lives the CFT, namely the $N=4$ super 
Yang--Mills theory with a {\it global} R-symmetry $SU(4)$ which 
is anomalous. In this case, however, the non-invariance of the 
5-dimensional Chern--Simons term does not provide an anomaly 
cancelling inflow, but explains the global $SU(4)$ anomaly of 
the CFT (see e.g. \cite{CB}). The mathematics is the same, 
but its interpretation 
is different. In brane world scenarios, on the other hand, anomaly 
cancellation may be a valuable constraint.

\section*{Acknowledgments}
This contribution is based largely on work done together with
Steffen \break\hfill
Metzger \cite{BM}. 

\vskip 1.cm


\end{document}